\documentclass[twocolumn]{aastex61}
\usepackage{epsfig,graphicx,graphics,subfigure}
\usepackage[T1]{fontenc}
 \usepackage{aecompl}
 \usepackage{natbib}

\begin{document}
\title{Alternative gravity rotation curves for the LITTLE THINGS Survey}
\correspondingauthor{James G. O'Brien}
\email{obrienj10@wit.edu, obrienj10@wit.edu}

\author{James ~G. ~O'Brien}
\affil{Wentworth Institute of Technology\\
550 Huntington Ave. \\
Boston, MA  02115}

\author{Thomas L.  Chiarelli}
\affil{Wentworth Institute of Technology\\
550 Huntington Ave. \\
Boston, MA  02115}

\author{Jeremy Dentico}
\affil{Wentworth Institute of Technology\\
550 Huntington Ave. \\
Boston, MA  02115}

\author{Modestas Stulge}
\affil{Wentworth Institute of Technology\\
550 Huntington Ave. \\
Boston, MA  02115}

\author{Brian Stefanski}
\affil{Wentworth Institute of Technology\\
550 Huntington Ave. \\
Boston, MA  02115}

\author{Robert Moss}
\affil{MIT Lincoln Laboratory\\
244 Wood St., Lexington , MA \\
02421, USA }

\author{Spasen Chaykov}
\affil{Brandeis University\\
15 South St. \\
Waltham MA 02453}

\begin{abstract}
Galactic rotation curves have proven to be the testing ground for dark matter bounds in spiral galaxies of all morphologies.  Dwarf galaxies serve as an increasingly interesting case of rotation curve dynamics due to their typically rising rotation curve as opposed to the flattening curve of large spirals.  Dwarf galaxies usually vary in galactic structure and mostly terminate at small radial distances.  This, coupled with the fact that Cold Dark Matter theories struggle with the universality of galactic rotation curves, allow for exclusive features of alternative gravitational models to be analyzed.  Recently, THINGS (The HI Nearby Galactic Survey) has been extended to include a sample of 25 dwarf galaxies now known as the LITTLE THINGS Survey.  Here, we show an application of alternative gravitational models to the LITTLE THINGS survey, specifically focusing on conformal gravity and Modified Newtonian Dynamics.  In this work, we provide an analysis and discussion of the rotation curve predictions of each theory to the sample.  Further, we show how these two alternative gravitational models account for the recently observed universal trends in centripetal accelerations in spiral galaxies.  This work highlights the similarities and differences of the predictions of the two theories in dwarf galaxies.  The sample is not large or diverse enough to strongly favor a single theory, but we posit that both conformal gravity and MOND can provide an accurate description of the galactic dynamics in the LITTLE THINGS sample without the need for dark matter.   
\end{abstract}


\section{Introduction}
For over seventy five years, the concept of dark matter has been widely accepted as the explanation for the missing mass problem in spiral galaxies. Many large scale collaborations have been created to search for dark matter with little success. In the current paradigm, the parameter space of cold dark matter is being constrained to higher and higher degrees, but physical searches have turned up with null or conflicting results  (see for example the recent Cryogenic Dark Matter Search (CDMS) (\cite{Agnese}). While the community further pushes technology to probe deeper into the possible parameter space of dark matter, the lack of pure observable evidence has lead others to the exploration of alternative theories to standard Einstein Gravity. Many of these alternative theories attempt to solve the rotation curve problem without invoking dark matter. Although many of these alternatives, such as Modified Newtonian Dynamics (MOND) (see \cite{MONDreview}), Scalar Tensor Vector Gravity (STVG) (see \cite{moffatt}), and more recently the Luminous Convolution Model (LCM) (see \cite{lcm}), have had success in fitting the rotation curves of spiral galaxies,  many of these theories struggle with creating a universal model of galactic rotational dynamics. A recent PRL publication (\cite{mcgaugh2016}) also furthers the idea that dark matter could be circumvented by new dynamical laws  only enhances the plausibility of alternative gravitational models.\\

For many years now, MOND has had significant success in fitting rotation curves of spiral galaxies of all morphologies (see for example, \cite{mcgaughfull}).  Since its founding by Milgrom, MOND's ability to capture rotation curve dynamics propelled the theory to the forefront of alternative gravitational models.  Recently, another alternative, conformal gravity (CG)  has fit a diverse set of over 130 rotation curves of various galaxies. \cite{fitting} have successfully fit the most recent data of the THINGS survey (\cite{Walter2008}) as well as the Ursa Major galaxies of  \cite{Verheijen1999}. Their research continued with less studied low surface brightness galaxies of \cite{Kim2007} and a survey of dwarf galaxies by  \cite{Swaters2002a}. To further diversify these studies, CG has been applied to three Tidal Dwarf Galaxies (TDG) of NGC 5291 with an astonishing degree of success (\cite{tdg}).  With these surveys fully populating the spectrum of spiral galaxy rotation curves, conformal gravity has emerged as an alternative gravitational theory that can universally explain the rotation curve problem.  

 Although extremely large galaxies serve as the perfect laboratory for new physics introduced in any competing alternative gravitational theory, small galaxies such as dwarfs provide a test case for universal applicability of a model.  
In this paper, we focus our efforts on the recent LITTLE THINGS survey of \cite{lTHINGS}, a survey of 25 dwarf galaxies offering ultra high resolution of rotation curve data.  In this work, we fit the rotation curves of the LITTLE THINGS data set with both CG and MOND and show that each theory can accommodate the rotation curve data without the need for dark matter.  A strong feature of this work is that this is  the first time that CG and MOND are fit simultaneously to particular galaxies using standardized input parameters to produce their respective predictions. In previous articles, these two theories would publish results separately, sometimes with conflicting input parameters, and this work is a step towards a more comprehensive comparison which can be used in future (and past) galaxy samples.

\section{ The LITTLE THINGS Survey}
Data for the Local Irregular That Trace Luminosity Extremes in The HI Nearby Galaxy Survey (LITTLE THINGS) was collected using the NRAO Very Large Array (VLA) located in New Mexico. The survey is a fairly local compilation, where each of the galaxies included are within 11 Mpc of the Milky Way. The observations of the galaxies were used to create rotation curves and were combined with the Spitzer archival 3.6$\mu$m and ancillary optical U, B, and V images to construct mass distribution models  (\cite{lTHINGS}). 
	From analyzing the mass distributions, the seminal paper addresses the cusp/core distribution problem of Cold Dark Matter (CDM) simulations, which predicts the luminous density slopes of galaxies.  
	

The paper also calls for more analysis on high-resolution low mass dwarf galaxies to further test the theory of CDM. For our purposes, we use the data collected to make mass model predictions for the rotation curves using MOND and CG, as well as identify trends in the centripetal accelerations of the sample for the two theories.

\section{Models of the rotation curves}
\subsection{General Relativity Prediction}
To provide an understanding of the missing mass problem in spiral galaxies mathematically, we present a brief review.  Starting from a spherical mass solution to the Einstein field equations (the Schwarzschild solution), a spiral galaxy is modeled as a distribution of point masses to produce an effective surface density given as:
\begin{equation}
\Sigma(R)=\Sigma_0e^{-\frac{R}{R_0}},\\
\label{sig}
\end{equation}
where {$R_0$} is the luminous scale length, and {$\Sigma_0$} is the central density. Upon integrating over the disk in cylindrical coordinates, one arrives at the familiar;

\begin{equation}
v_{gr}(R) = \sqrt{\frac{N^*\beta^*c^2R^2}{2R^3_0}F_b(R)},\\ 
\label{gr}
\end{equation}
where
\begin{equation}
F_b = \left[I_0\left(\frac{R}{2R_0}\right)K_0\left(\frac{R}{2R_0}\right)-I_1\left(\frac{R}{2R_0}\right)K_1\left(\frac{R}{2R_0}\right)\right],
\end{equation}
such that $I_0$, $I_1$, $K_0$, and $K_1$ are Bessel functions and {$\beta^*=\frac{M_\odot G}{c^2}=1.48*10^5$} cm.  This is the well established Freeman curve, where the only free parameter is the overall mass of the target galaxy, 
\begin{equation}
N^* = \frac{M_{disk}}{M_{\odot}}.
\end{equation}
Although {$N^*$} is a free parameter for fitting purposes, it is physically bounded by the preservation of the mass to light ratio to be on the order of unity.   It has been long established (\cite{Rubin1980}) that the  prediction of eq. (\ref{gr}) typically does not match the observed data.  This effect becomes even more pronounced outside the inner region of a galaxy (typically at $R>2 R_0$), and the mass discrepancy shows in the rising rotation curves of dwarfs, and the flattening rotation curves of large spirals.

\subsection{Gas and bulge contributions}
\label{gmasses}
Since the dwarf galaxies contain a non-negligible gas component, all of the above fitting predictions must take into account the gas contributions.  For each, the mass of the gas (listed in Table \ref{t1}) is taken directly from \cite{lTHINGS}, which already includes the contributions of He.  Using this mass in eq. (\ref{gr}) with its respective gas scale length allows us to account for the gas contribution.  Typically, one would also model a spherical bulge when needed.
However, for the current LITTLE THINGS sample, there are no documented bulges (\cite{lTHINGS}), resulting in the net rotation curve consisting of contribution by gas and stars alone,
\begin{equation}
v_{NEW}(R) = \sqrt{v^2_{gr}(R)+v^2_{gas}(R)}.
\label{vnew}
\end{equation}

\subsection{Cold Dark Matter (CDM) models}
One method of solving the missing mass problem is to increase the mass of the galaxy outside the $R>2 R_0$ region typically fit by eq. (\ref{gr}). However, since this mass is not observed in the optical or radio bands, we assume it to be dark.  In order to make the data match the prediction of eq. (\ref{gr}), we can assume the total rotational velocity would be modified as 
\begin{equation}
v_{total}(R) = \sqrt{v^2_{NEW}(R) + v^2_{dark}(R)},
\end{equation}
where $v_{dark}$ is the contribution to the rotational velocity from the  missing mass.
Using a non singular isothermal Newtonian spherical distribution, $\sigma(R)=\frac{\sigma_0}{(1+(\frac{r}{r_0})^2)} $ in hydrostatic equilibrium, (see for example \cite{mannheimfull}) the dark matter contribution can take the form:

\begin{equation}
v^2_{dark}(R) = 4\pi \beta^* c^2\sigma_0 r_0^2\left[1-\frac{r_0}{R}arctan\left(\frac{R}{r_0}\right)\right],
\label{dark}
\end{equation}
where $\sigma_0$ is the spherical dark matter density and $r_0$ is the dark matter core radius.  The function described in eq. (\ref{dark}) is not the only choice for the distribution of the dark matter in a galaxy.  Other forms have been shown to reproduce the observed rotation curve data (see for example, \cite{mannheimfull}), but still retain the same two parameters $\sigma_0$ and $r_0$ as in eq. (\ref{dark}).  Since we have not yet physically observed dark matter, these two parameters have to be fit to the data against  the total observed velocity $v_{obs}$ as measured by astronomers. The two free parameters effectively shape the modeled rotation curve,  and when coupled with the free parameter of the total luminous mass, $M_{disk}$, the theory has three total free parameters. It should be noted that although this is only two extra free parameters per individual galaxy, this process must be arbitrarily done for any studied galaxy, making the number of fitting parameters for a given sample three times the number of galaxies studied. Further, due to the nature of eq. (\ref{dark}), the dark matter contribution will force the rotation curves to become asymptotically flat.  In \cite{impact}, it is shown that large galaxies can show departures from asymptotic flatness, making large spirals a great test case for the comparison of dark matter and alternative gravity models. Although this feature is not  accessible in the LITTLE THINGS survey, CG and MOND with fixed parameters can be applied to the rising rotation curves of these dwarfs.  This allows us to highlight similarities and differences in the modeling of each and will be explored in Section \ref{diffs}.

\subsection{Conformal gravity formulation}
Conformal gravity, originally derived by Weyl, was later re-studied by Mannheim and Kazanas \cite{Mannheim1989}. Conformal gravity retains a completely covariant metric theory of gravity but also includes the feature of local conformal invariance. The action is unchanged due to local transformations $g_{\mu\nu}(x)\rightarrow e^{2\alpha(x)}g_{\mu\nu}(x)$ with local phase $\alpha(x)$.  Since CG is still a metric theory of gravity, many of the familiar properties of General Relativity such as curvature of space and time,  the coupling of electromagnetic fields to gravity (\cite{sultana2010}), and the precession of the orbit of Mercury (\cite{sultana}), are retained.  While CG was not originally studied to solve the rotation curve problem, one may use the rotation curve problem as a testing ground for the overall theory. Conformal gravity assumes a scalar action which is described by

\begin{equation}
I_{\rm W}=-\alpha_g\int d^4x (-g)^{1/2}C_{\lambda\mu\nu\kappa}
C^{\lambda\mu\nu\kappa} \\,
\label{182}
\end{equation}
where \\
\begin{eqnarray}
C_{\lambda\mu\nu\kappa}&= &R_{\lambda\mu\nu\kappa}
-\frac{1}{2}\left(g_{\lambda\nu}R_{\mu\kappa}\right) \nonumber \\
& &
+\frac{1}{2}\left(g_{\lambda\kappa}R_{\mu\nu}-
g_{\mu\nu}R_{\lambda\kappa}+
g_{\mu\kappa}R_{\lambda\nu}\right) \nonumber \\
& &
+\frac{1}{6}R^{\alpha}_{\phantom{\alpha}\alpha}\left(
g_{\lambda\nu}g_{\mu\kappa}-
g_{\lambda\kappa}g_{\mu\nu}\right),
\label{180}
\end{eqnarray}
and $C_{\lambda\mu\nu\kappa}$ is the conformal Weyl Tensor.  The resulting gravitational field equations for conformal gravity can be recovered in the usual prescription (see for example \cite{wburg}) by varying eq. (\ref{182}) with respect to the metric tensor, yielding the field equations as
\begin{equation}
4\alpha_g W^{\mu\nu}=4\alpha_g\left[
2C^{\mu\lambda\nu\kappa}_
{\phantom{\mu\lambda\nu\kappa};\lambda;\kappa}-
C^{\mu\lambda\nu\kappa}R_{\lambda\kappa}\right]
 =T^{\mu\nu}.
\label{188}
\end{equation}
%

%

Since the goal  is to obtain the conformal gravity prediction for a rotation curve,  one must first begin with the conformal gravity equivalent of a ``Schwarzschild like" solution. We can recover the typical vacuum solution to eq. (\ref{188}) since $W^{\mu\nu}$ vanishes when $R^{\mu\nu}$ vanishes. However, Mannheim and Kazanas showed that the vanishing solutions may not be the only solutions to the fourth order theory. They found, as described in \cite{Mannheim1989}, that in the conformal theory the exact line element is given by $ds^2=-B(r)dt^2+B(r)^{-1}dr^2+r^2d\Omega_2$, where the exterior metric coefficient $B(R>a)$ is given by 
\begin{equation}
B(R>a)=1-\frac{2\beta}{R}+\gamma R -kR^2,
\label{E1}
\end{equation}
outside a massive gravitational source of radius a.
We can immediately see the departures from the Einstein formalism, wherein the two new factors $\gamma$ and $k$ arise due to the fact that the conformal theory is fourth order. It should also be noted that when $\gamma$ and $k$ are small in eq. (\ref{E1}), we return the exact Schwarzschild solution of standard gravity.  We can now  follow the standard procedure  by describing a galaxy as disk with exponential density falloff in the radial direction as in eq. (\ref{sig}). Upon using the solution in eq. (\ref{E1}), one obtains the rotational velocity prediction for a disk of masses as

\begin{equation}
v_{cg}(R) = \sqrt{{v_{gr}^2+\frac{N^*\gamma^*c^2R^2}{2R_0}F_{\gamma^*}(R)}{+\frac{\gamma_0c^2R}{2}-\kappa c^2R^2}},
\label{total}
\end{equation}
where the integration constants are set as
\begin{eqnarray}
\gamma^* & = & 5.42*10^{-41} {\rm cm}^{-1},\nonumber \\
\gamma_0 & = & 3.06*10^{-30} {\rm cm}^{-1},  \nonumber\\
\kappa & = & 9.54*10^{-54}~{\rm cm}^{-2}, \nonumber
\end{eqnarray} 
and
\begin{equation}
F_{\gamma^*}(R)=I_1\left(\frac{R}{2R_0}\right)K_1\left(\frac{R}{2R_0}\right). \nonumber
\end{equation}
For a thorough discussion on the relevant integrations which set these constants, we refer the reader to \cite{fitting}.  Recently, in \cite{sultana}, it was also shown that accurate tests of the motion of Mercury's orbit can also provide bounds on these constants.
As can be seen in eq. (\ref{E1}) and (\ref{total}), the presence of the linear and quadratic potential terms are negligible on solar system scales, but would begin to dominate at large galactic scales.  Hence, CG admits a  universal formula which can then be tested without reference to the particular morphology of a given galaxy.  In order to include the gas contribution, we use the gas masses in Table  \ref{t1} in eq. (\ref{total}) as described in section \ref{gmasses}.  Typical CG fitting would also include other features such as a bulge as described in \cite{fitting}, but since the LITTLE THINGS survey contains no such bulges, they will be left out for this study.

\subsection{MOND formulation}

Since its construction, MOND has been the leading alternative gravitational theory to explain the missing mass problem in rotation curves.
Unlike conformal gravity, MOND directly modifies existing Newtonian dynamics through the rescaling of the gravitational acceleration via the following function: 	
		\begin{equation}
		g = \frac{g_{N}}{\mu(g/a_{0})},
		\end{equation}
		where {$g_{N}$} is the Newtonian gravitational acceleration, and $a_{0} \approx 1.21 \times 10^{-8} cm\,s^{-2}$ for a choice of interpolation function as given below.
	
The constant, $a_{0}$ is described broadly as the critical acceleration regime below which Newtonian gravity is not valid, and $\mu(x)$ is an asymptotic function (the interpolation function) such that $\mu(x) = 1$ when $g \gg a_{0}$ and $\mu(x) = g/a_{0}$ when $g \ll a_{0}$. 
	The original interpolation function given in eq. (\ref{MondInt1}) was shown to be a valid approximation for asymmetric discs (\cite{Gentile_2008}), and is given by
		\begin{equation}
		\mu_{orig}(x)=\frac{x}{\sqrt{1+x^2}}.
		\label{MondInt1}
		\end{equation}
However, the  MOND model can be implemented using varying asymptotic interpolation functions, as described in \cite{mcgaughint}.	A second interpolation function leads to a different prediction than the one  given by eq. (\ref{MondEq2}), and a slightly larger critical acceleration ($a_{0} = 1.35 \times 10^{-8} ~cm\,s^{-2}$), which resulted in better fits for the modeled dwarf galaxies, and has the form of:
		\begin{equation}
		\mu_{FB}(x)=\frac{x}{1+x}.
		\label{MondInt2}
		\end{equation}
		Since this paper is focusing on the dwarf galaxies, we shall assume the interpolation function of eq. (\ref{MondInt2}) for fitting purposes.  Using  eq. (\ref{MondInt2}), we get the total observed MOND velocity prediction in terms of the baryonic masses as:		
		\begin{equation}
		V^2_{MON}(R)=	V^2_{NEW}(R)+	V^2_{NEW}(R)\frac{\sqrt{1+\frac{4a_{0}R}{V^2_{NEW}(R)}}-1}{2}.
		\label{MondEq2}
		\end{equation}
It should be noted that proper MOND fitting takes place on a point by point analysis for photometric and HI data, which Gentile showed (\cite{Gentile_2008}) that eq. (\ref{MondEq2}) captures the essence of the full MOND analysis.  
In our fitting models, we included the Modified Newtonian Dynamics predictions of \cite{Gentile_2008} since eq. (\ref{MondEq2}) best models dwarfs, as a means to simultaneously compare CG and MOND on a per galaxy basis.

\section{The Fits}

\begin{figure*}
\epsfig{file=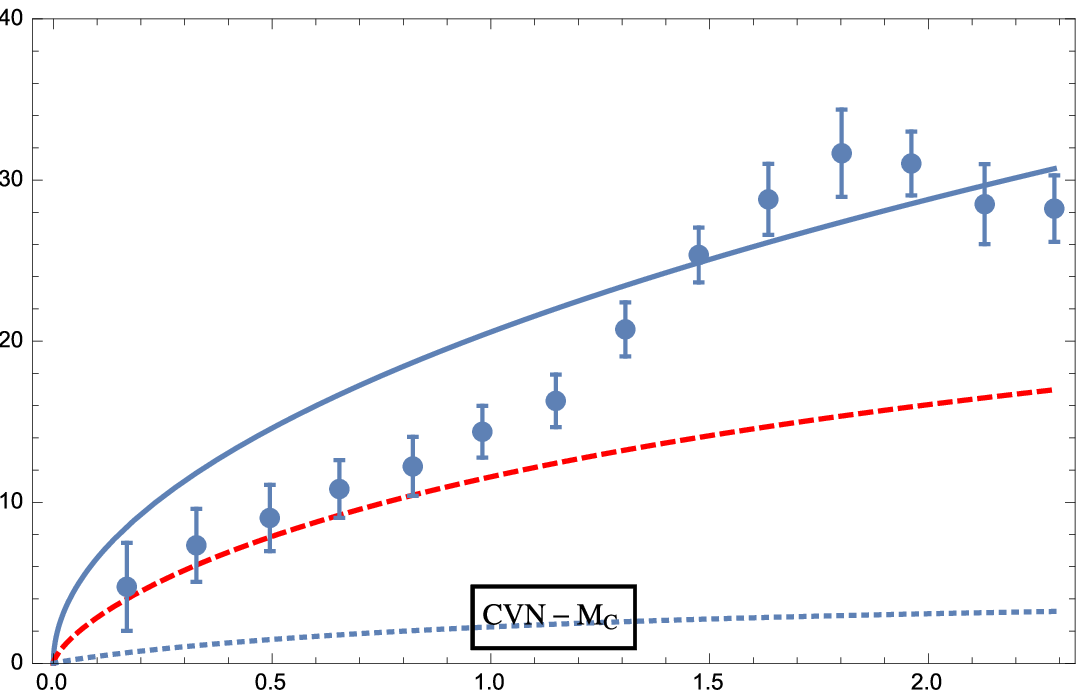,width=1.5825in,height=1.2in}~~~
\epsfig{file=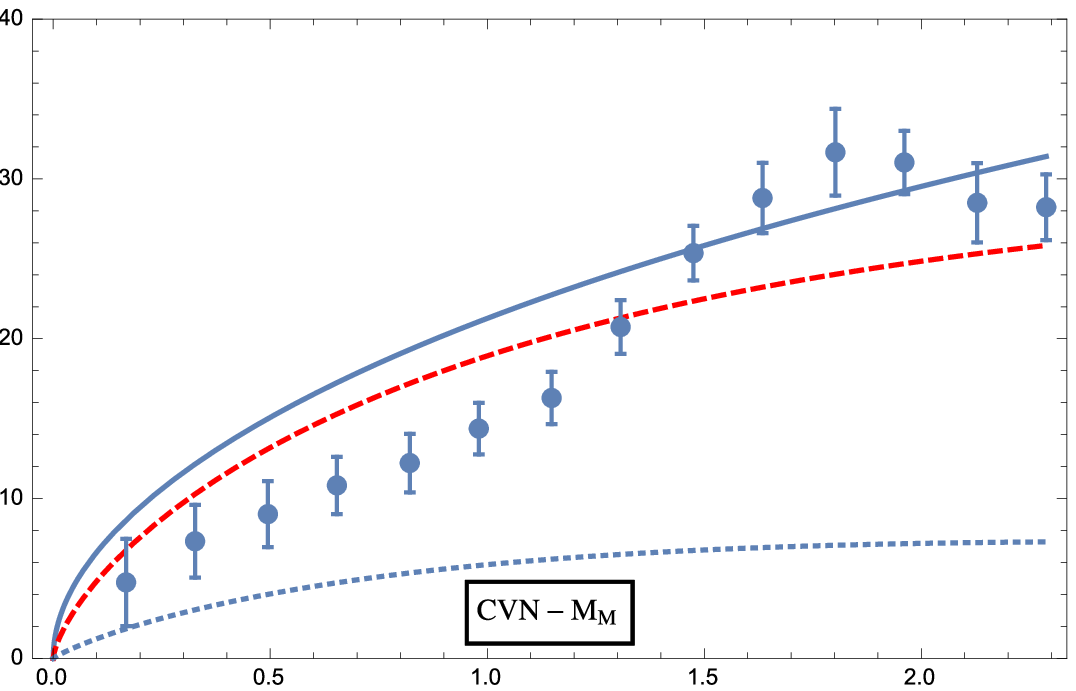,width=1.5825in,height=1.2in}~~~
\epsfig{file=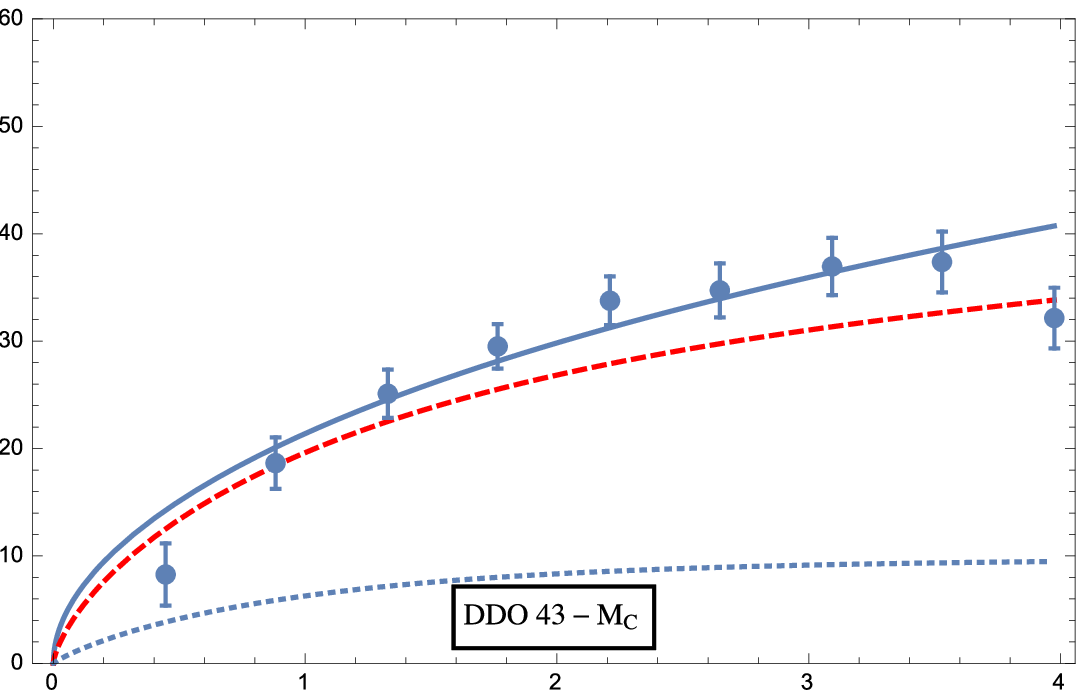,width=1.5825in,height=1.2in}~~~
\epsfig{file=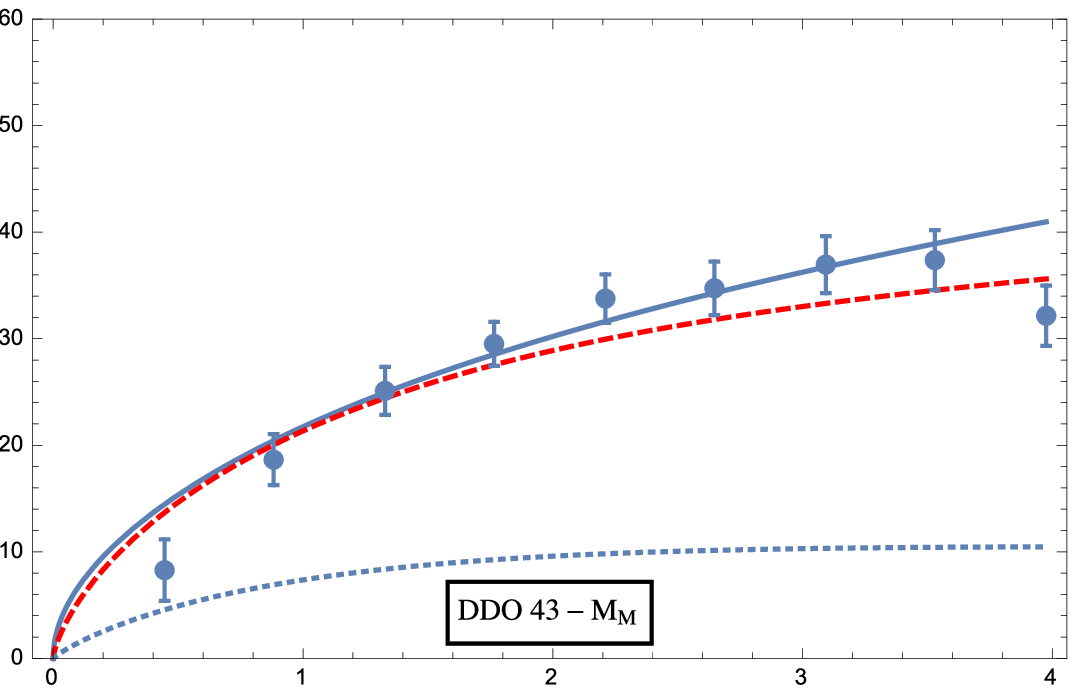,width=1.5825in,height=1.2in}\\
\smallskip
\epsfig{file=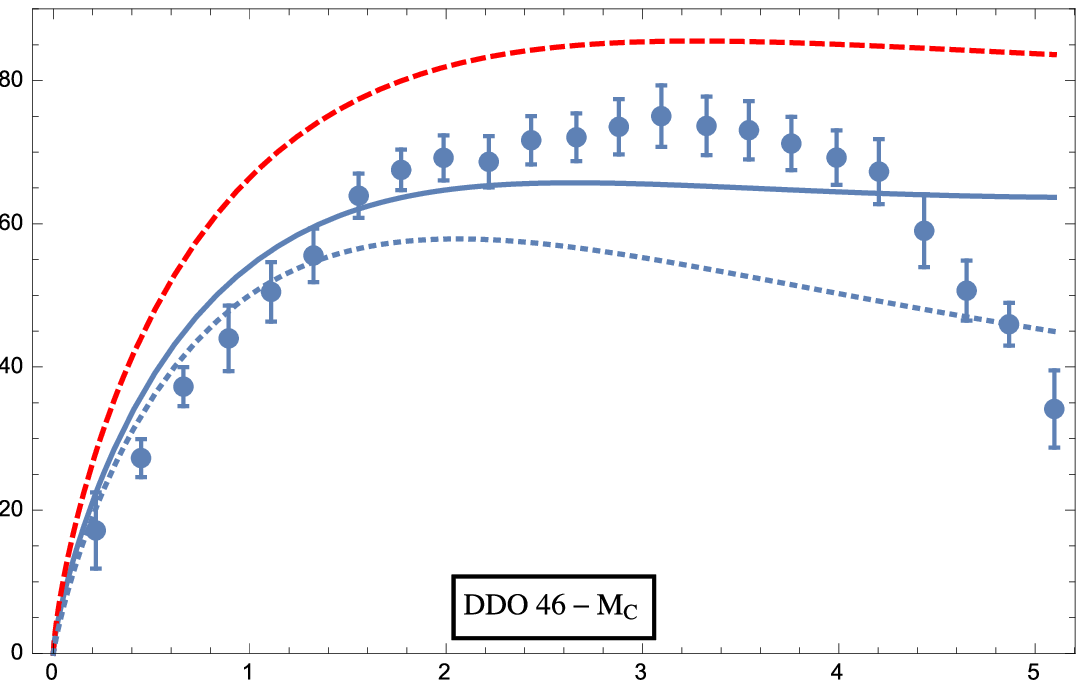,width=1.5825in,height=1.2in}~~~
\epsfig{file=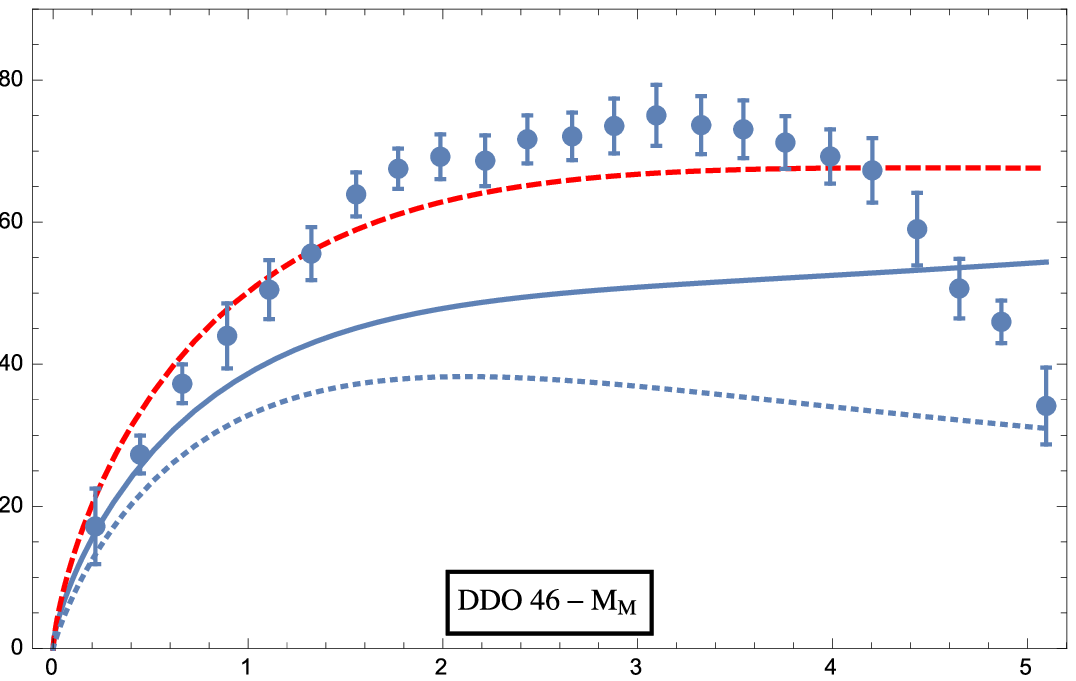,width=1.5825in,height=1.2in}~~~
\epsfig{file=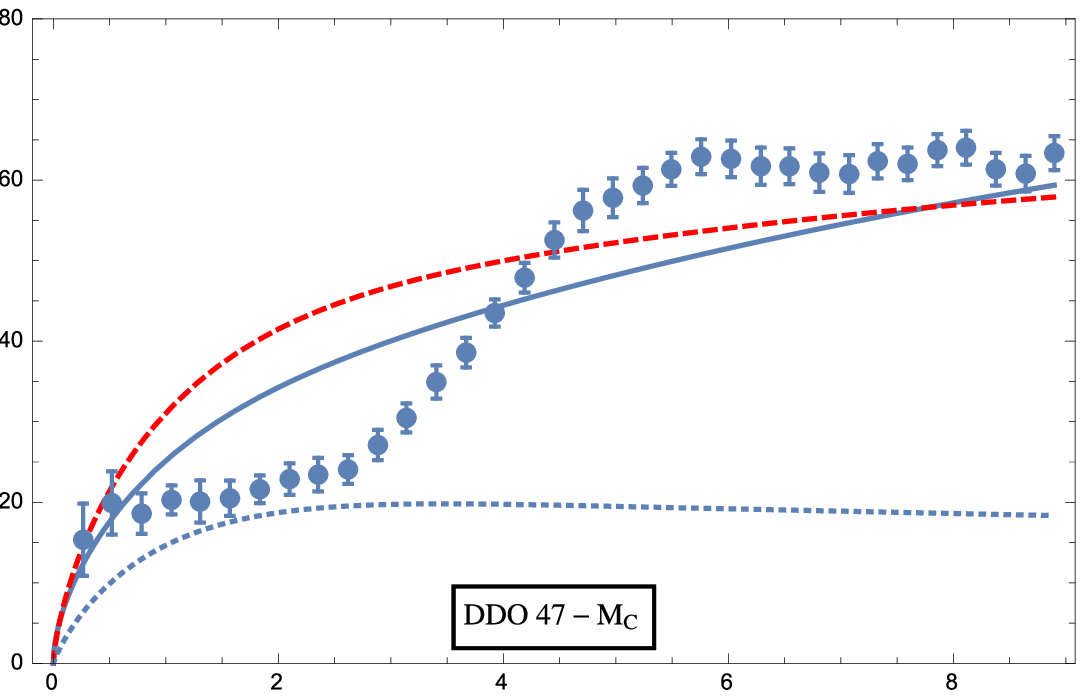,width=1.5825in,height=1.2in}~~~
\epsfig{file=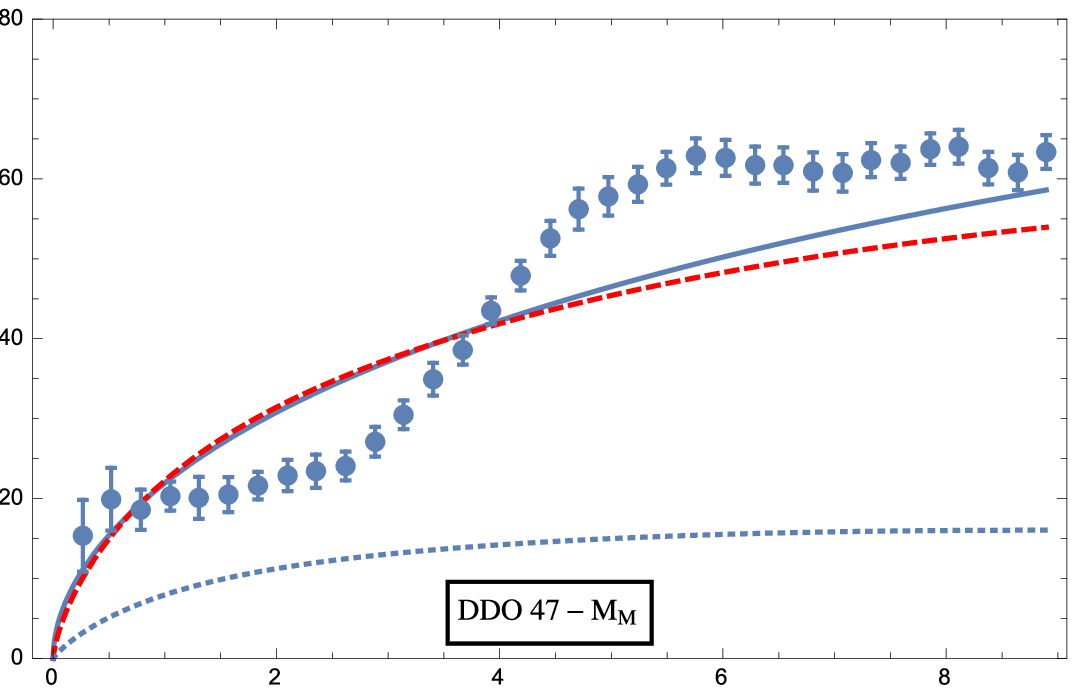,width=1.5825in,height=1.2in}\\
\smallskip
\epsfig{file=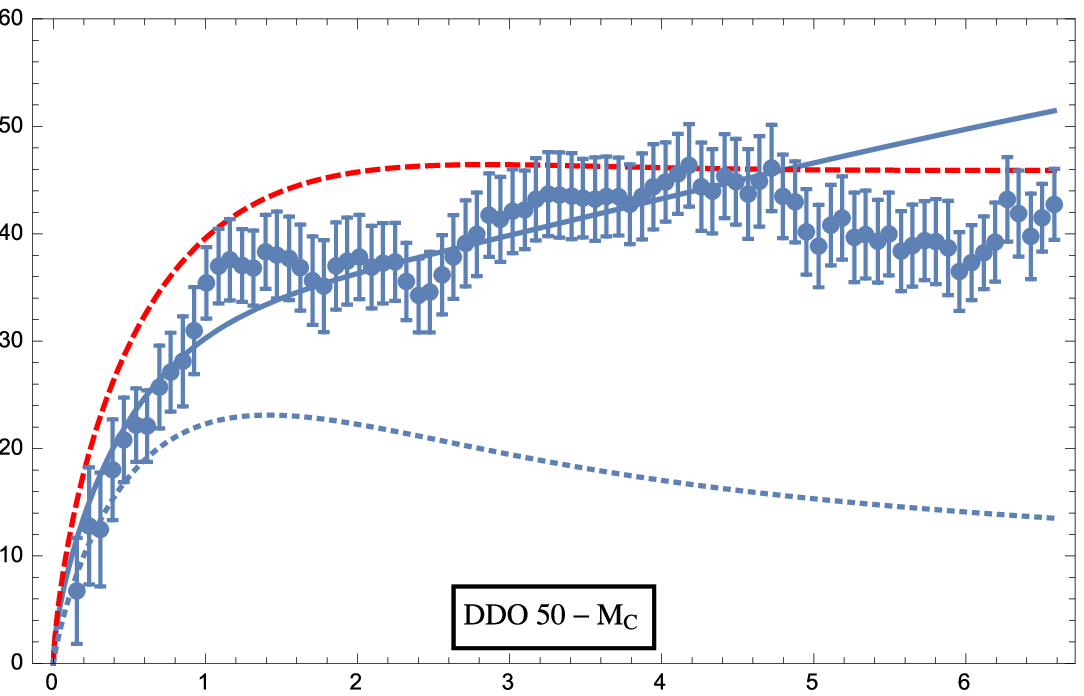,width=1.5825in,height=1.2in}~~~
\epsfig{file=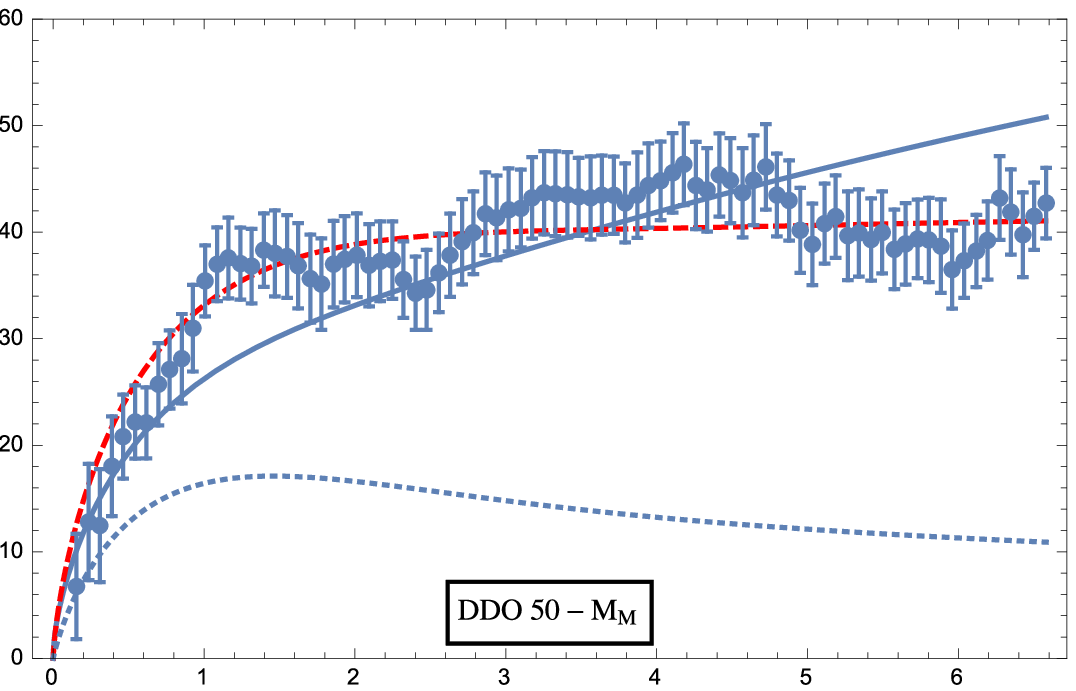,width=1.5825in,height=1.2in}~~~
\epsfig{file=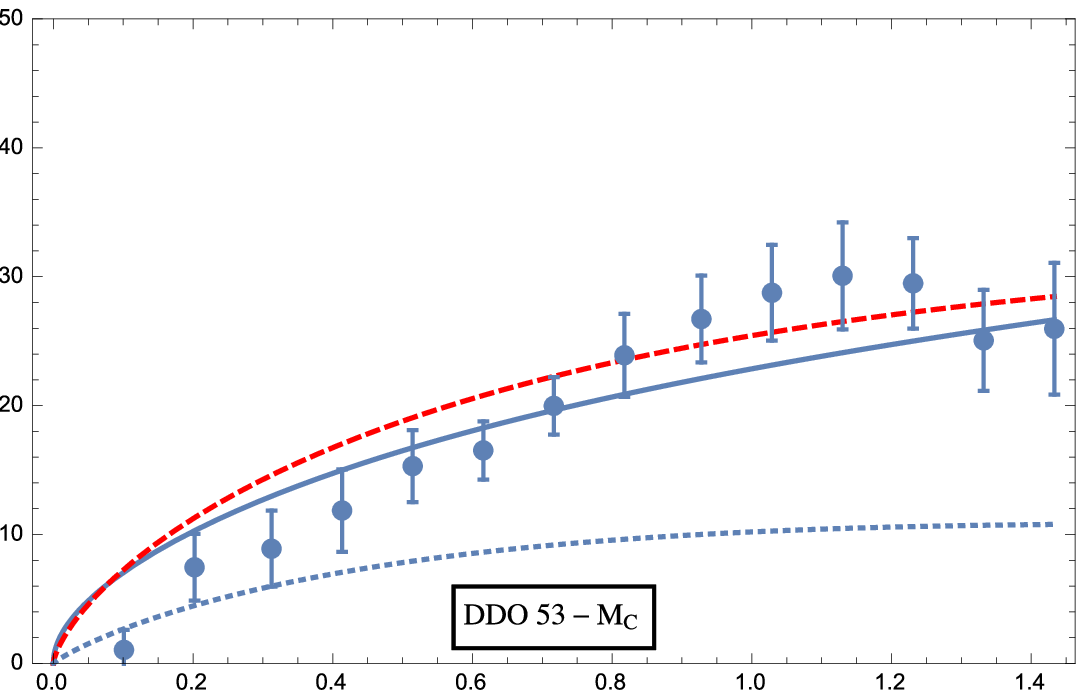,width=1.5825in,height=1.2in}~~~
\epsfig{file=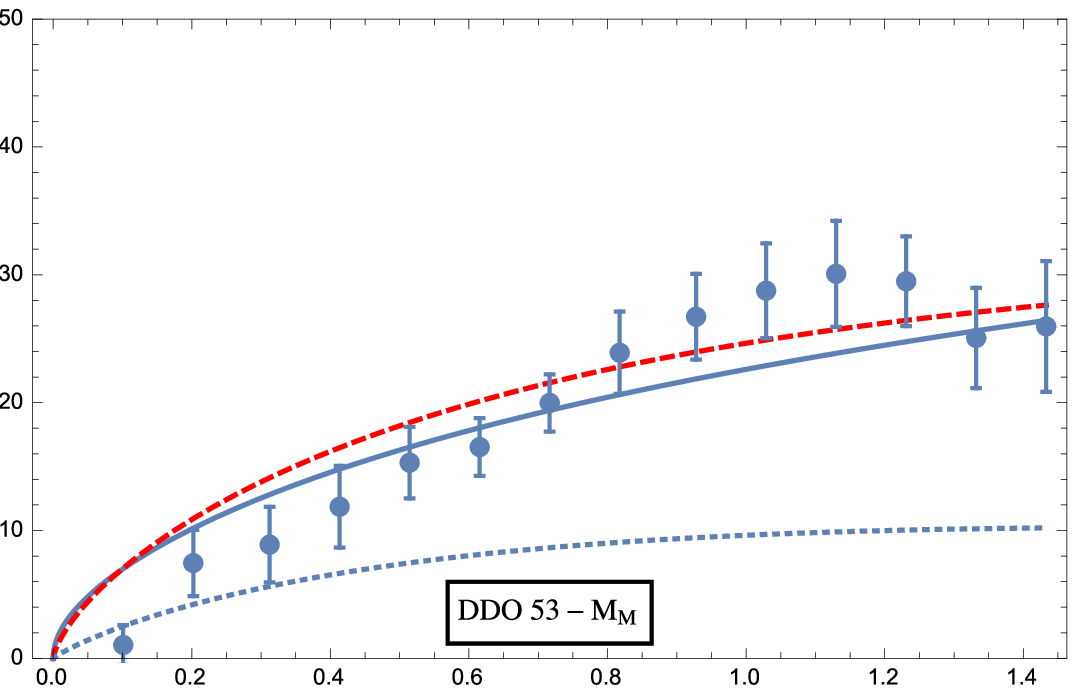,width=1.5825in,height=1.2in}\\
\smallskip
\epsfig{file=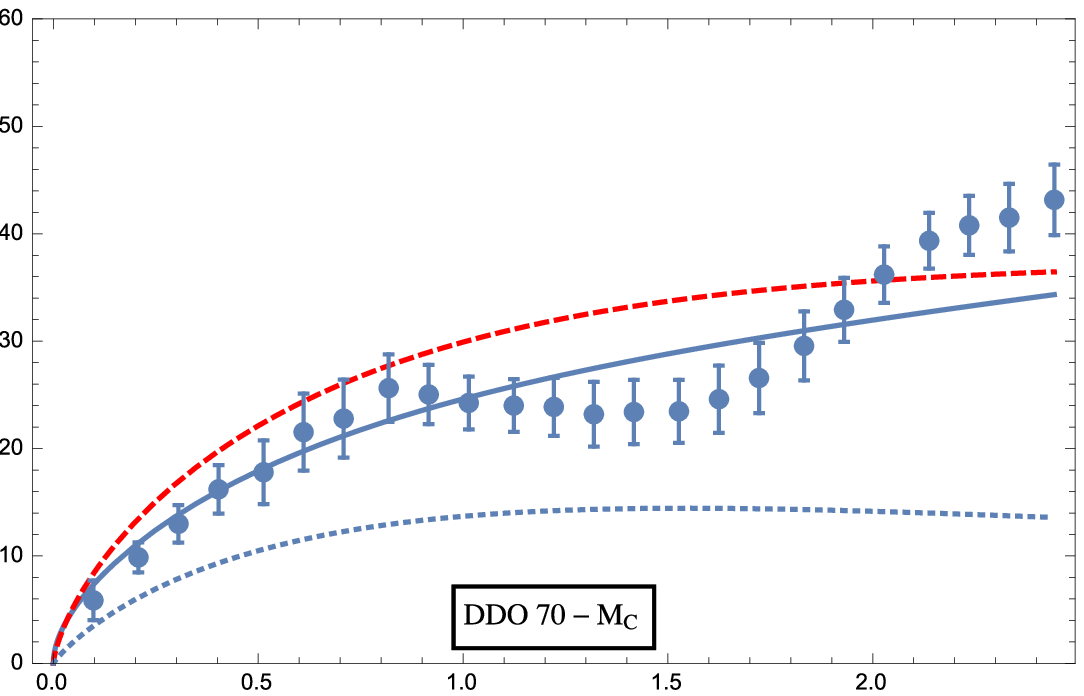,width=1.5825in,height=1.2in}~~~
\epsfig{file=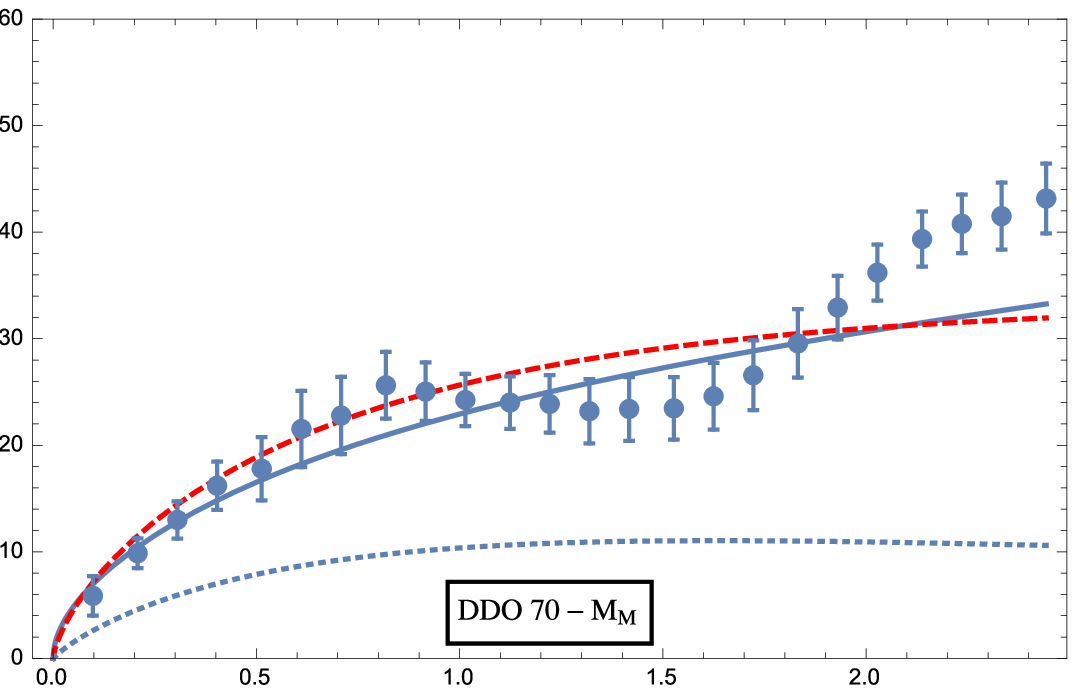,width=1.5825in,height=1.2in}~~~
\epsfig{file=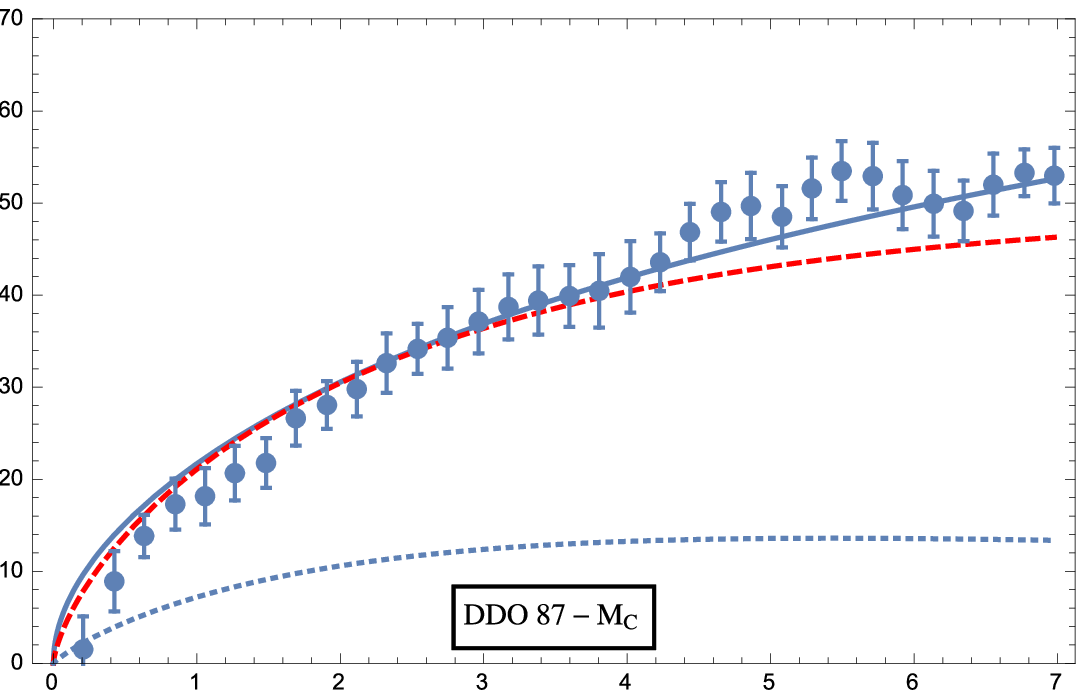,width=1.5825in,height=1.2in}~~~
\epsfig{file=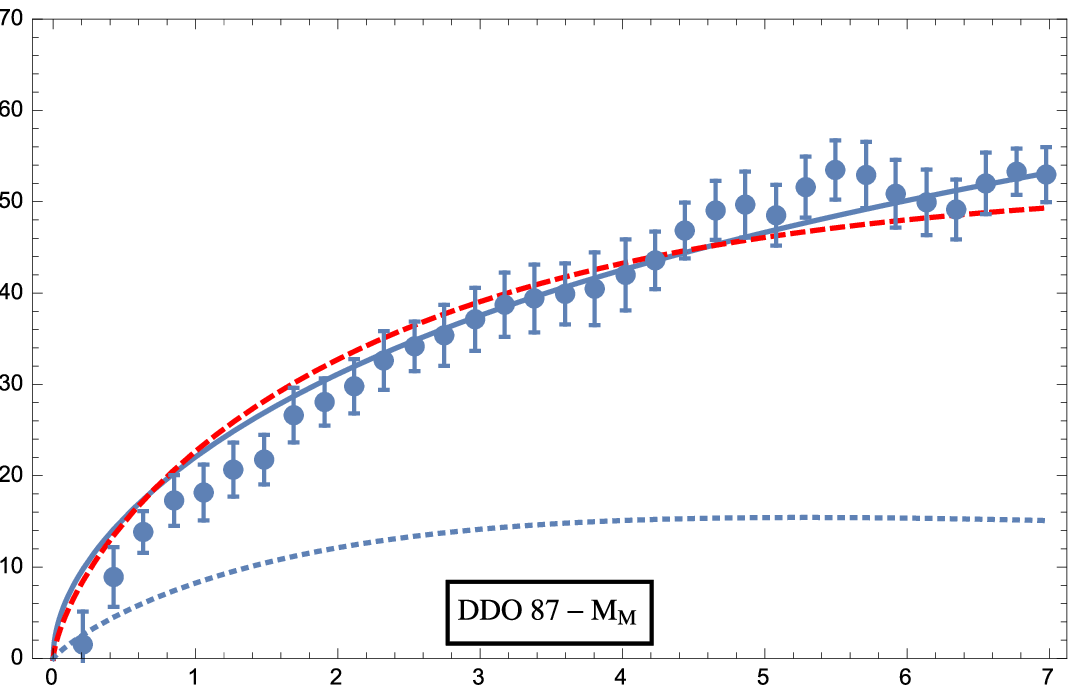,width=1.5825in,height=1.2in}\\
\medskip
\caption{Fitting to the rotational velocities (in ${\rm km}~{\rm sec}^{-1}$) of  the LITTLE THINGS  galaxy sample with their quoted errors as plotted as a function of radial distance (in ${\rm kpc}$). For each galaxy we have exhibited the contribution due to the Newtonian term alone (dashed curve), the full blue curve showing the conformal gravity prediction and the full red curve showing the MOND prediction. No dark matter is assumed.}
\label{rot1}
\end{figure*}

\begin{figure*}
\epsfig{file=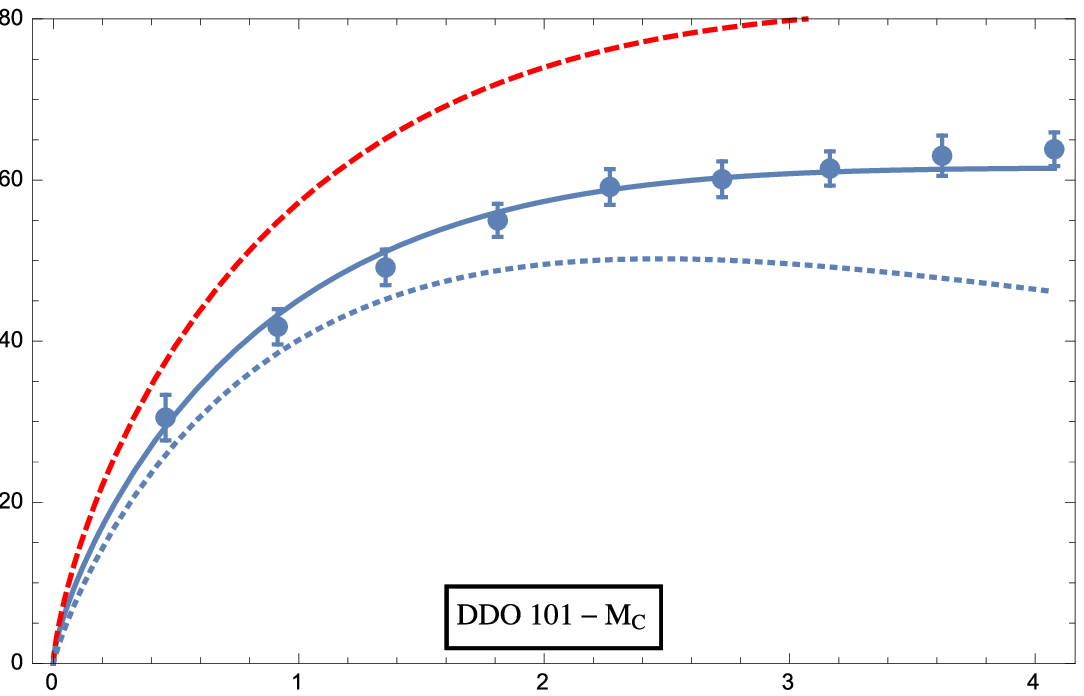,width=1.5825in,height=1.2in}~~~
\epsfig{file=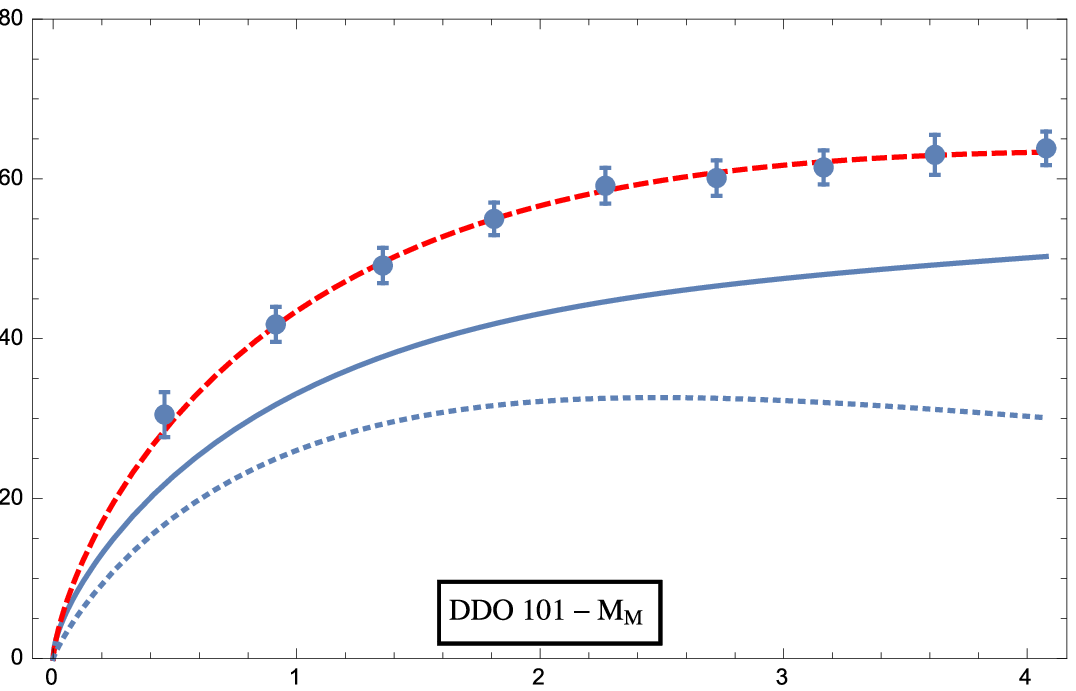,width=1.5825in,height=1.2in}~~~
\epsfig{file=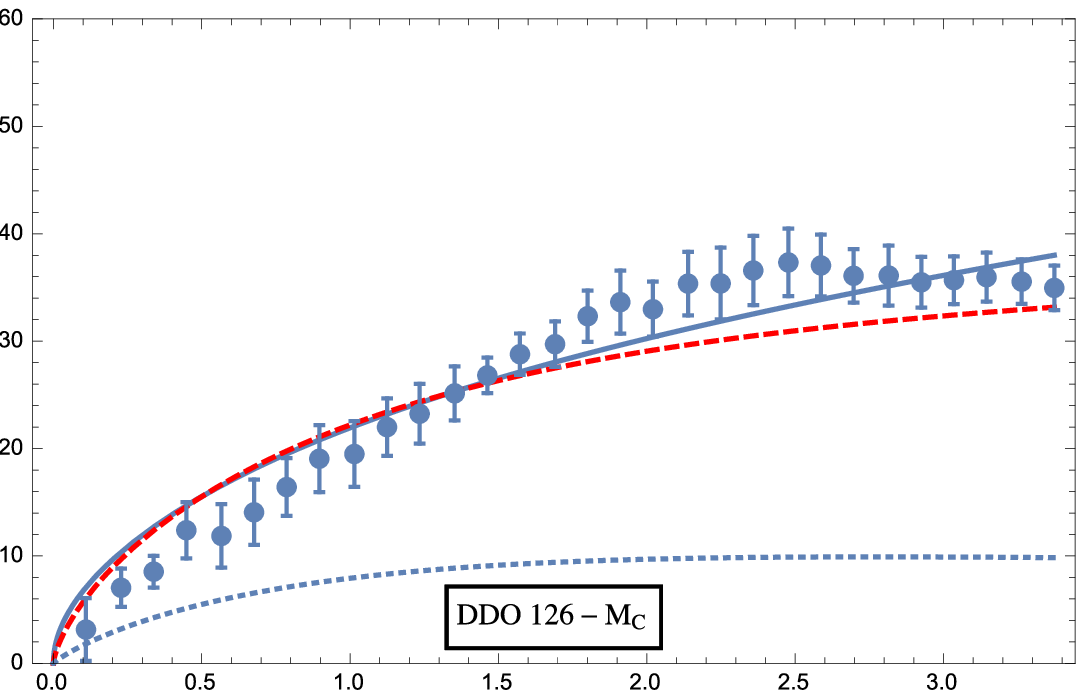,width=1.5825in,height=1.2in}~~~
\epsfig{file=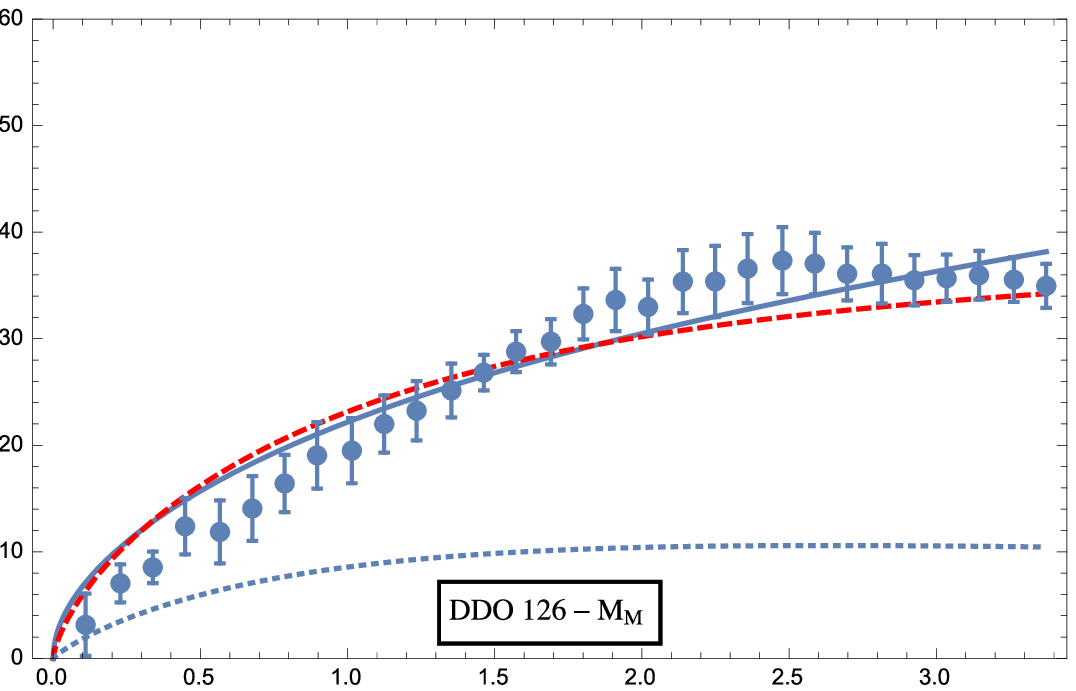,width=1.5825in,height=1.2in}\\
\smallskip
\epsfig{file=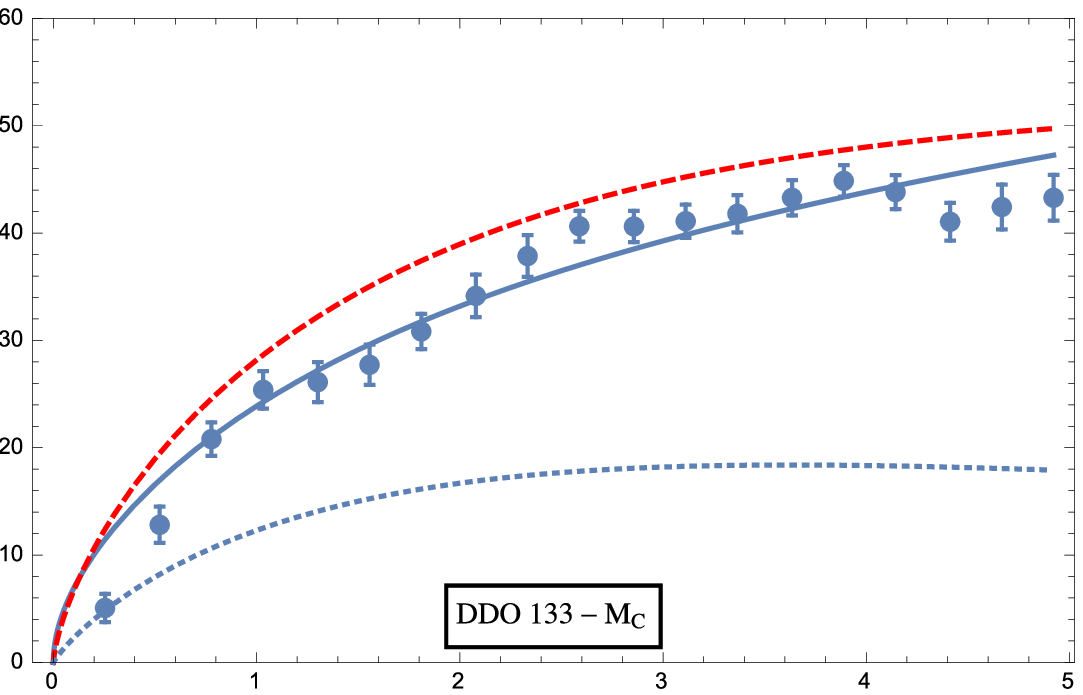,width=1.5825in,height=1.2in}~~~
\epsfig{file=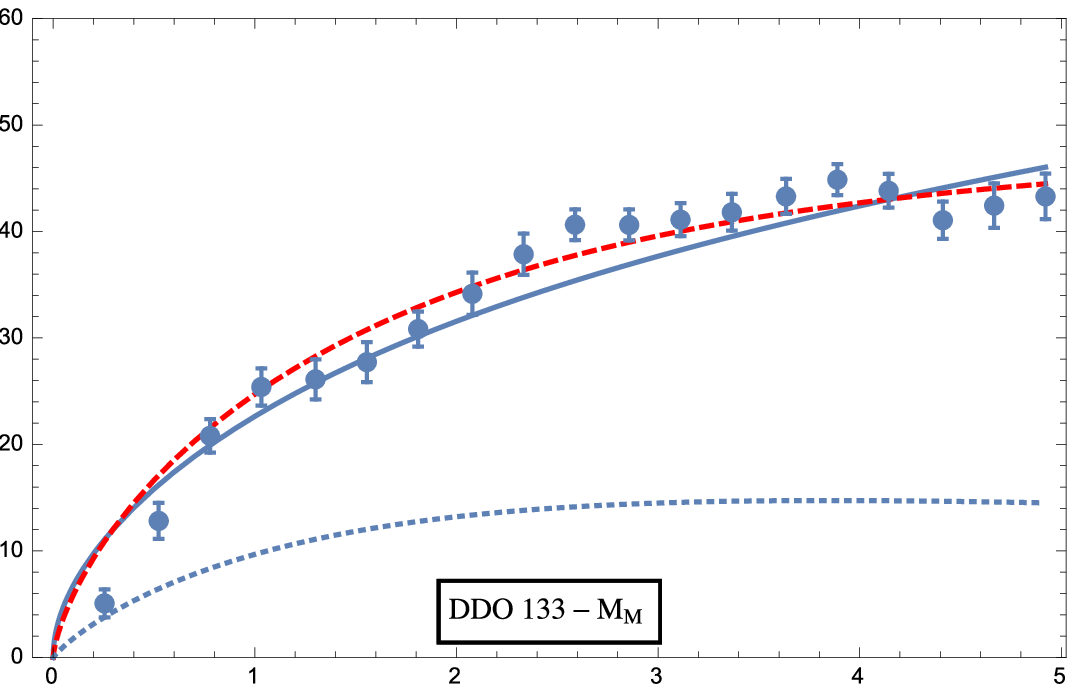,width=1.5825in,height=1.2in}~~~
\epsfig{file=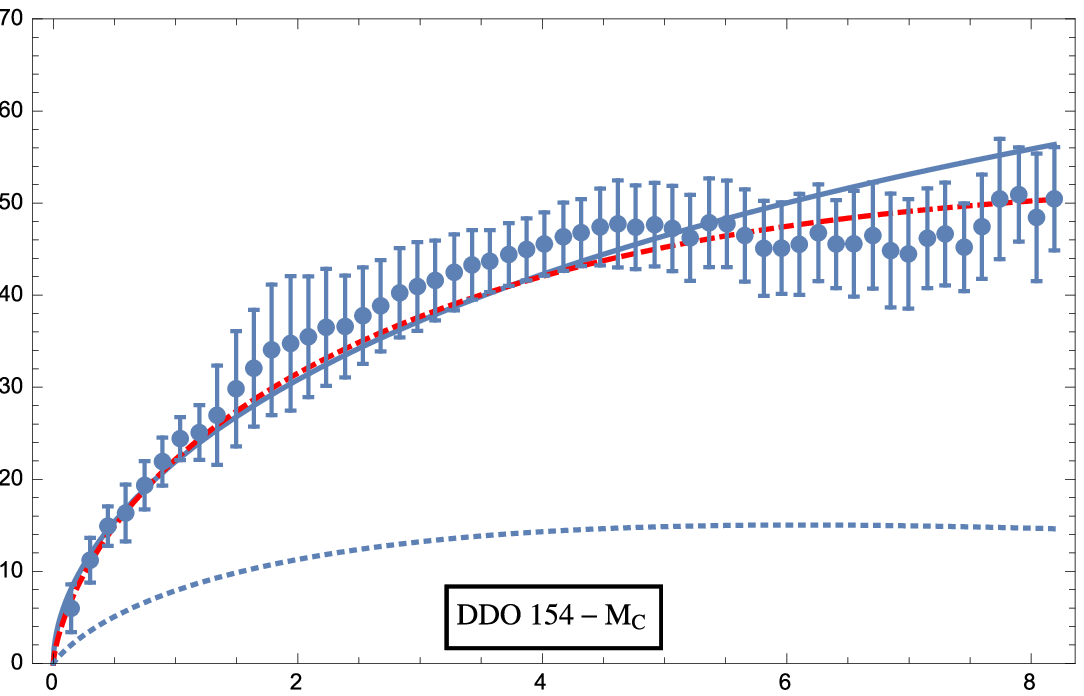,width=1.5825in,height=1.2in}~~~
\epsfig{file=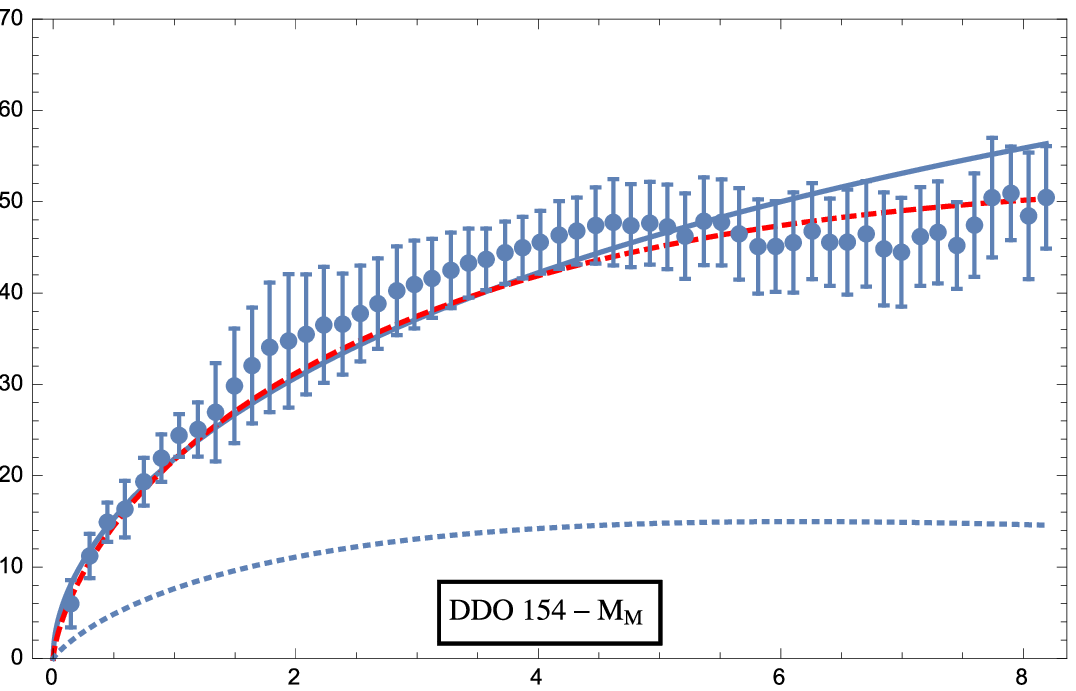,width=1.5825in,height=1.2in}\\
\smallskip
\epsfig{file=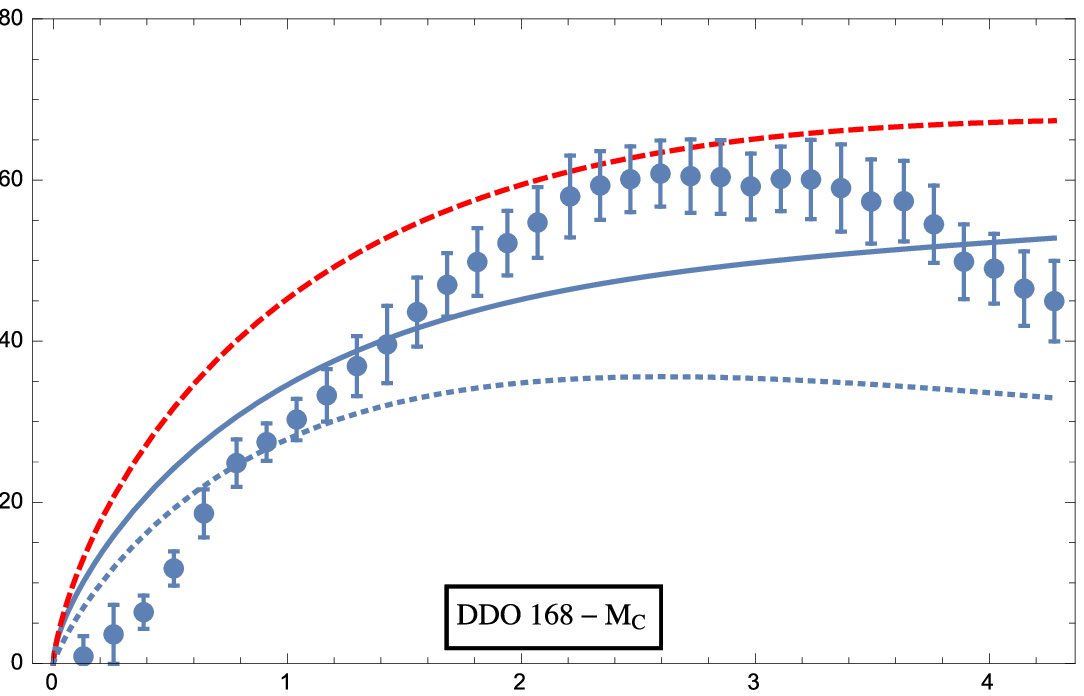,width=1.5825in,height=1.2in}~~~
\epsfig{file=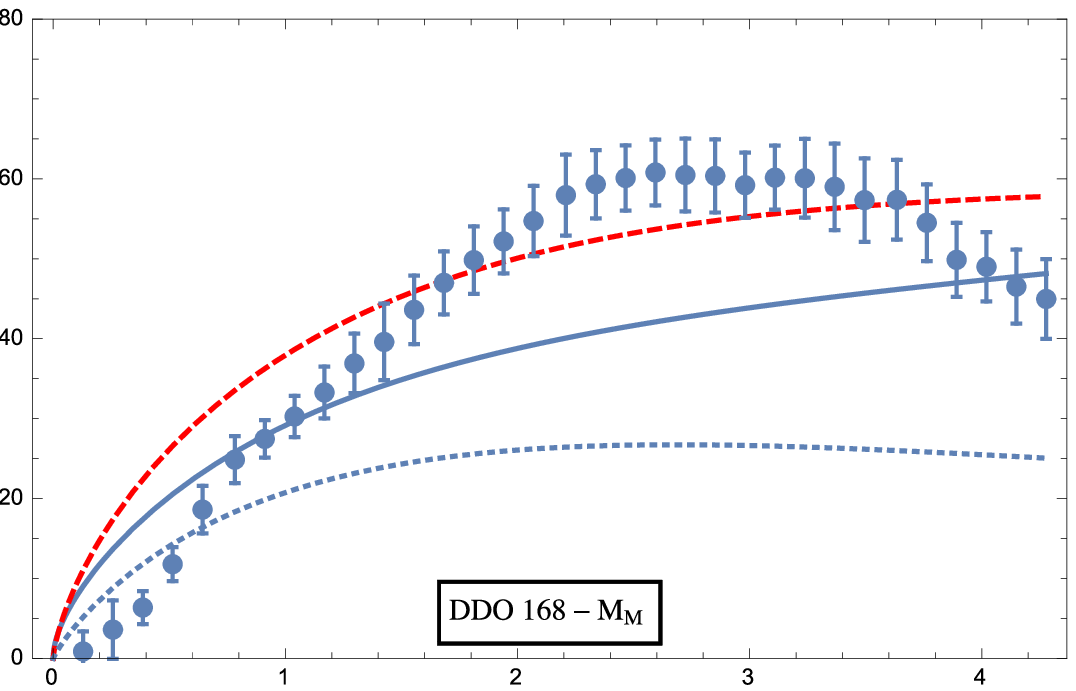,width=1.5825in,height=1.2in}~~~
\epsfig{file=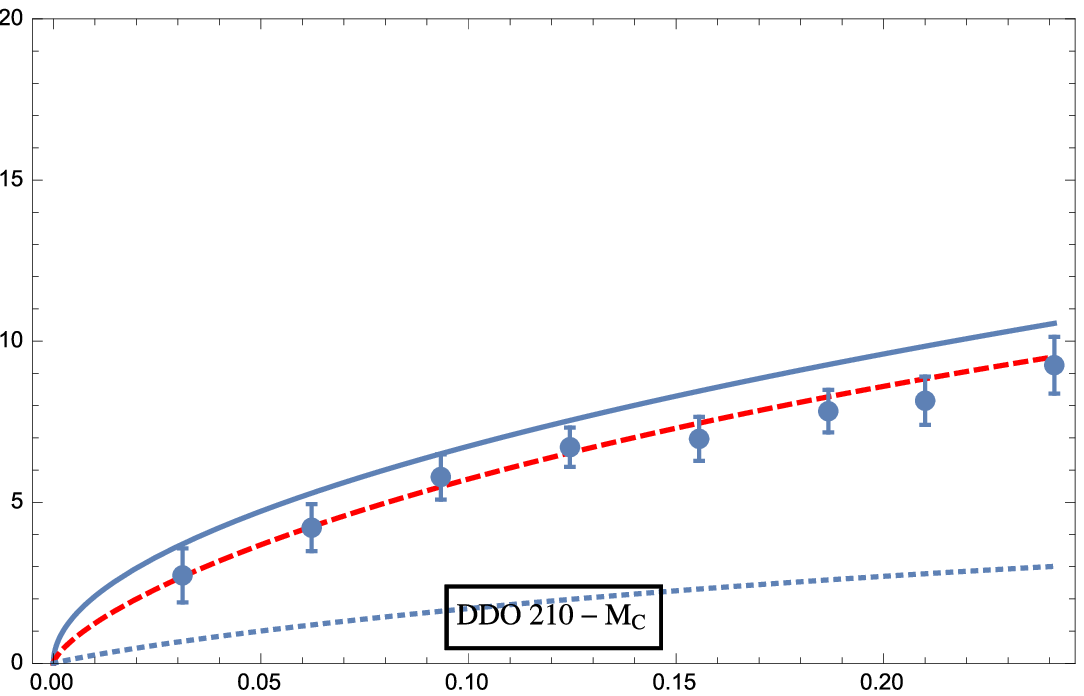,width=1.5825in,height=1.2in}~~~
\epsfig{file=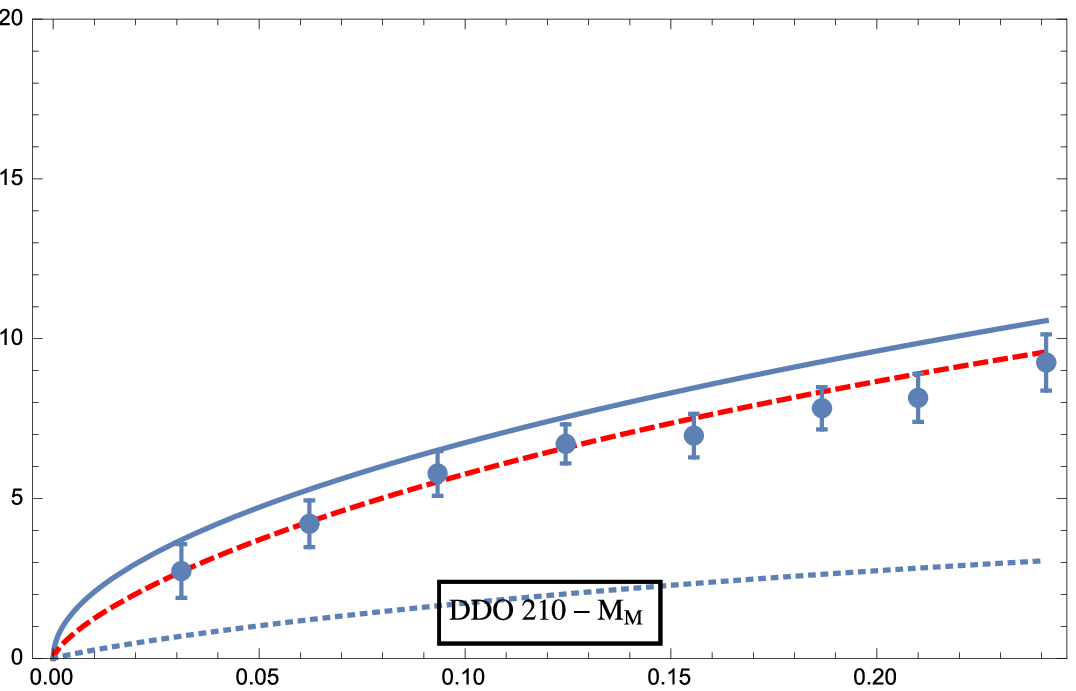,width=1.5825in,height=1.2in}\\
\smallskip
\epsfig{file=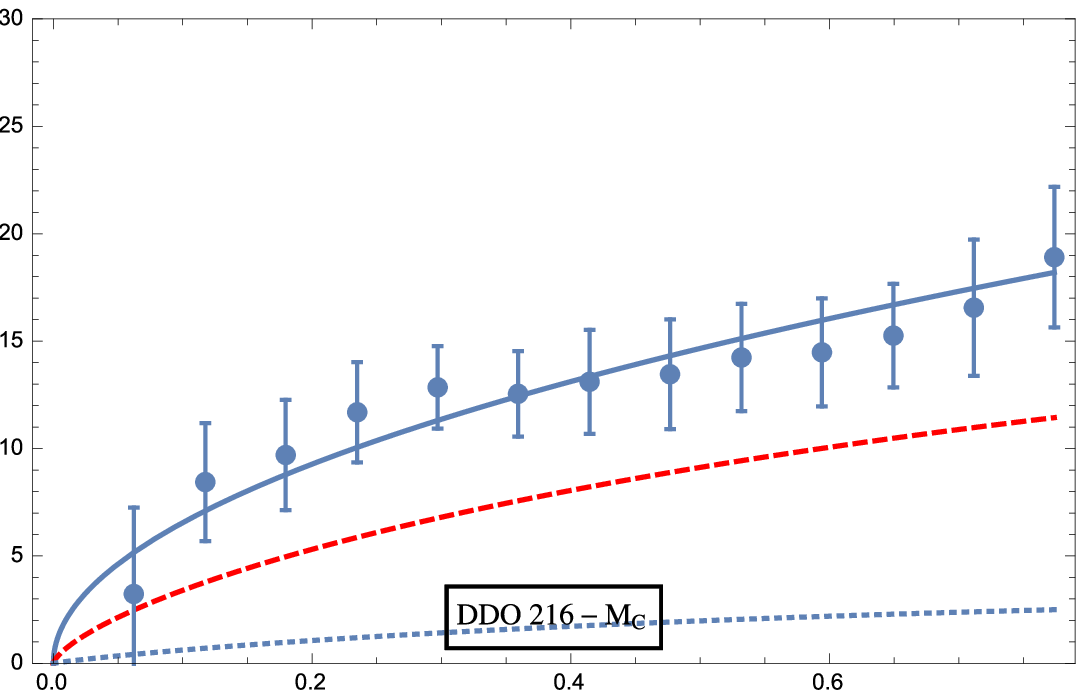,width=1.5825in,height=1.2in}~~~
\epsfig{file=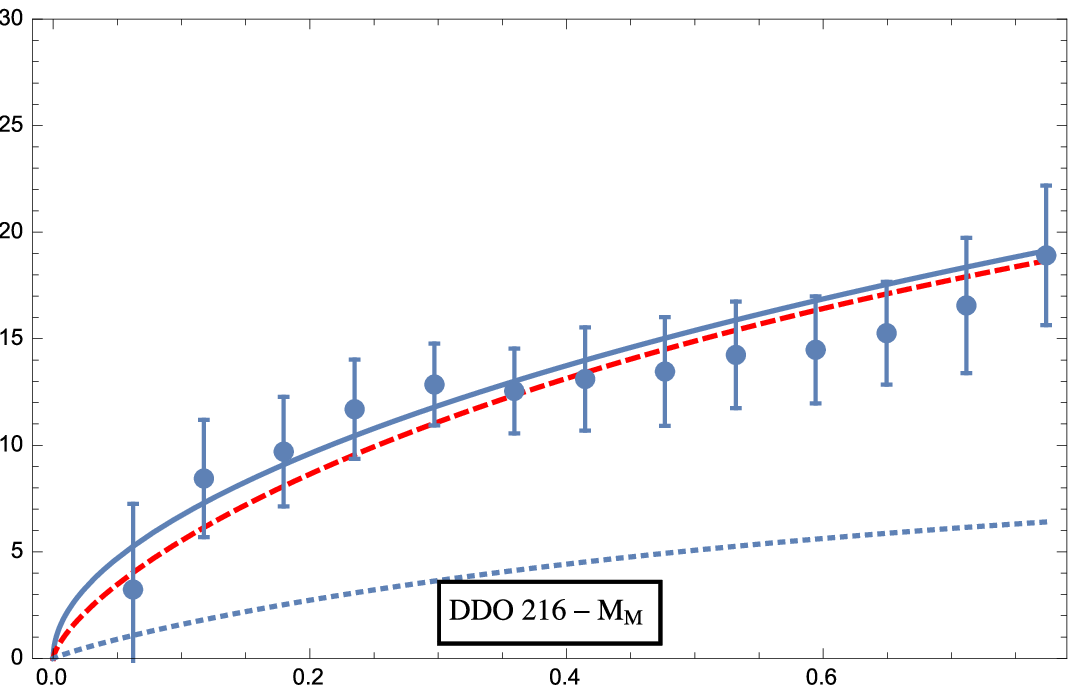,width=1.5825in,height=1.2in}~~~
\epsfig{file=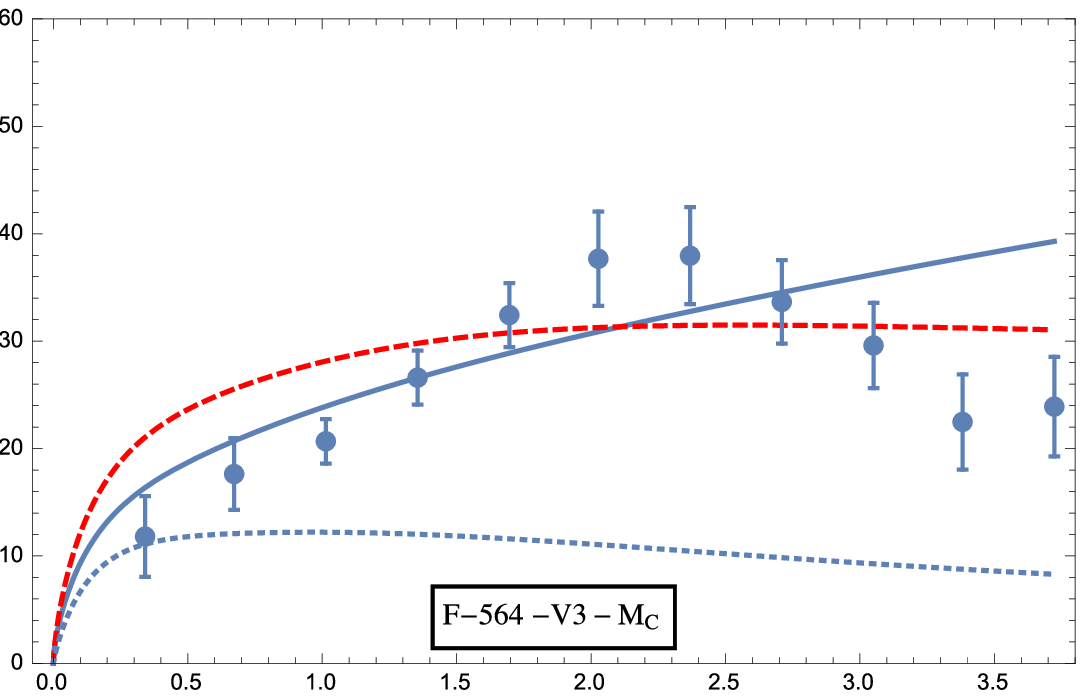,width=1.5825in,height=1.2in}~~~
\epsfig{file=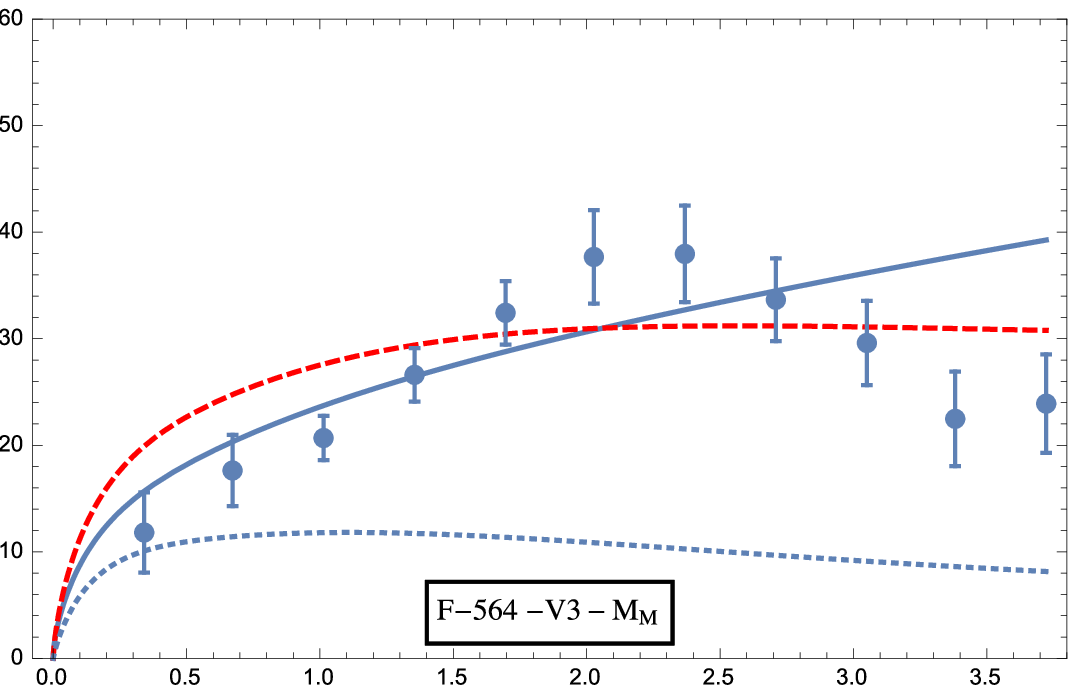,width=1.5825in,height=1.2in}\\
\medskip
\caption{Fitting to the rotational velocities (in ${\rm km}~{\rm sec}^{-1}$) of  the LITTLE THINGS  galaxy sample with their quoted errors as plotted as a function of radial distance (in ${\rm kpc}$). For each galaxy we have exhibited the contribution due to the Newtonian term alone (dashed curve), the full blue curve showing the conformal gravity prediction and the full red curve showing the MOND prediction. No dark matter is assumed.}
\label{rot2}
\end{figure*}

\begin{figure*}
\epsfig{file=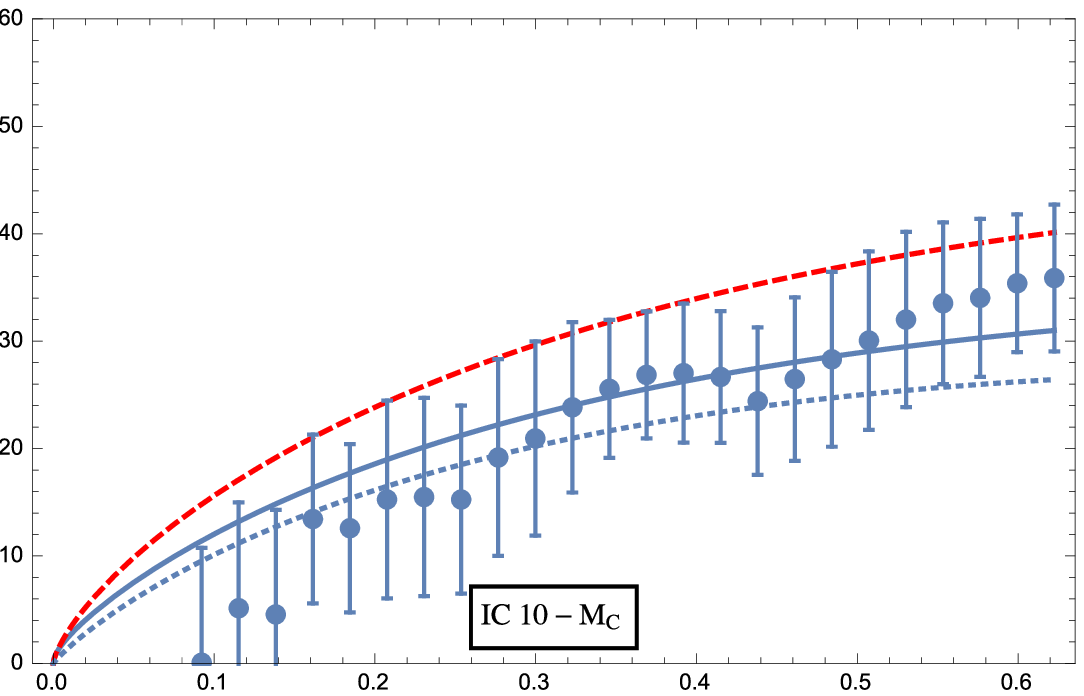,width=1.5825in,height=1.2in}~~~
\epsfig{file=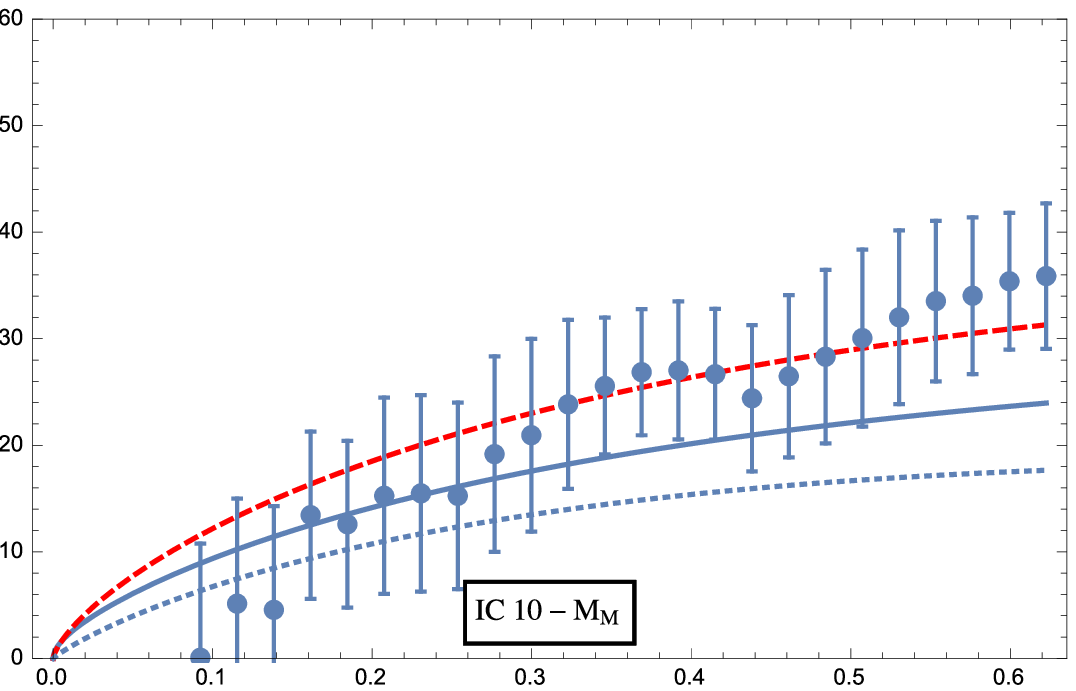,width=1.5825in,height=1.2in}~~~
\epsfig{file=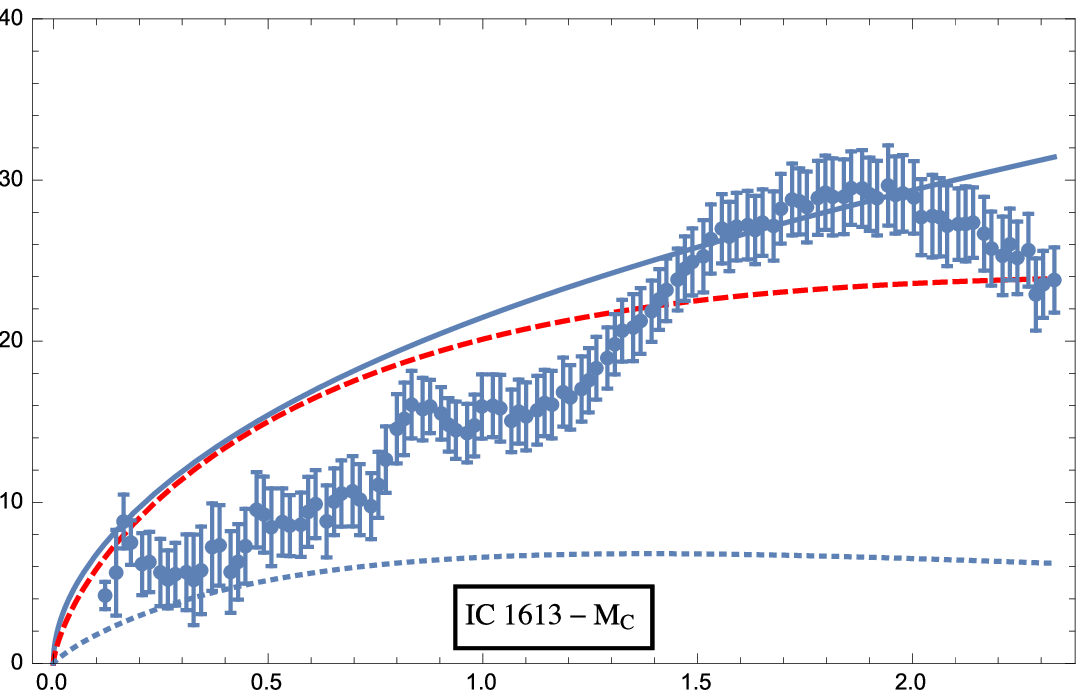,width=1.5825in,height=1.2in}~~~
\epsfig{file=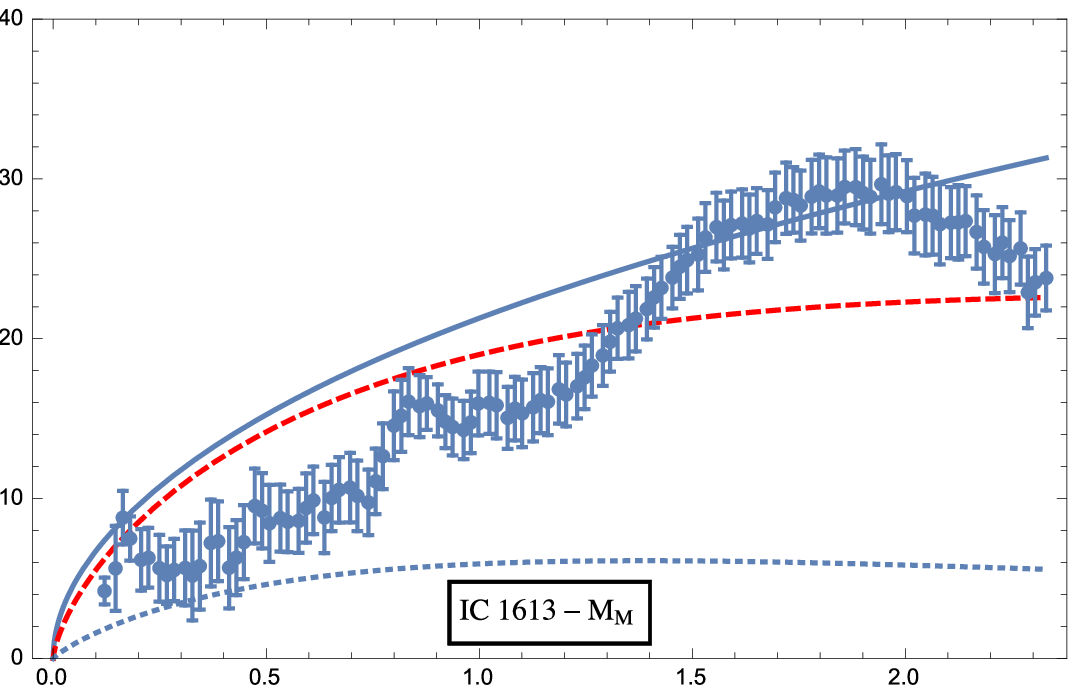,width=1.5825in,height=1.2in}\\
\smallskip
\epsfig{file=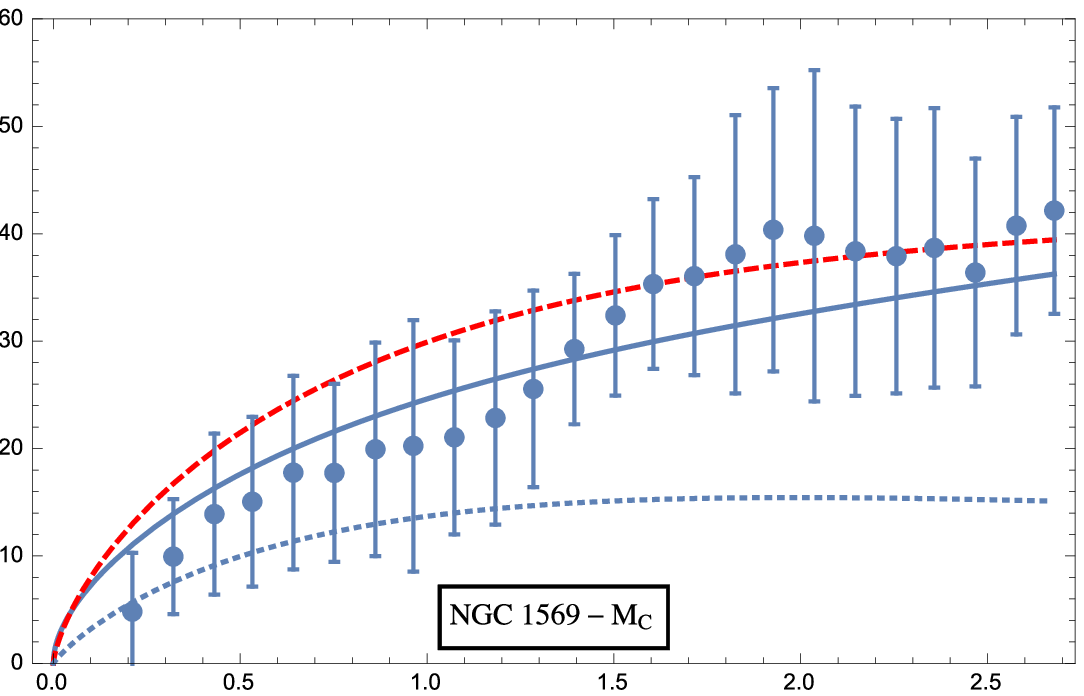,width=1.5825in,height=1.2in}~~~
\epsfig{file=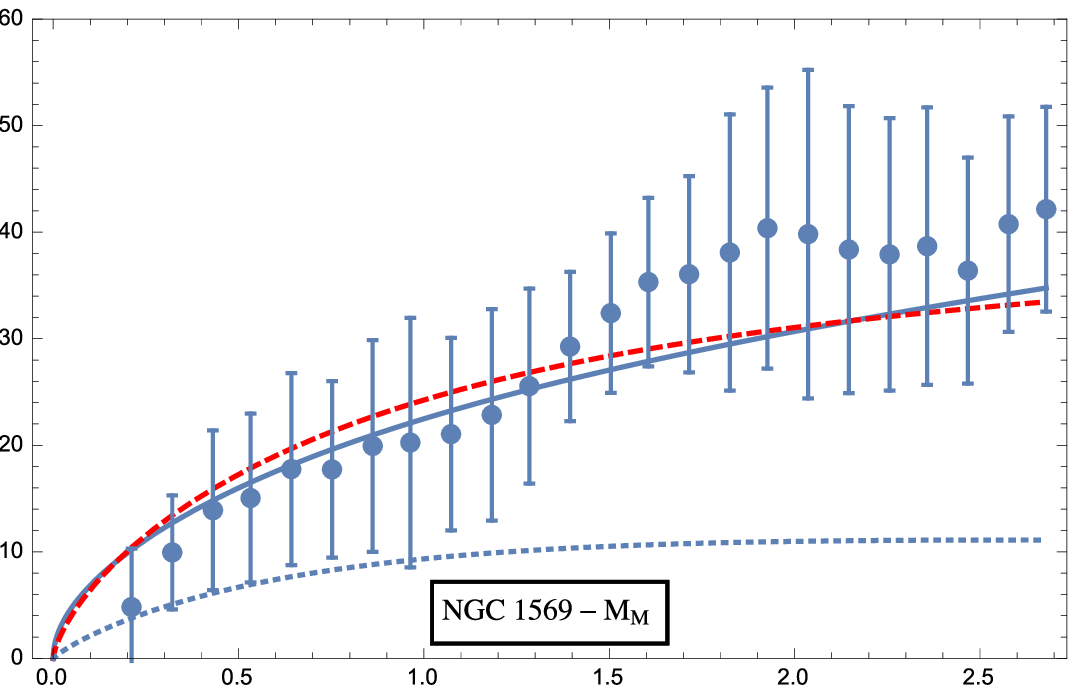,width=1.5825in,height=1.2in}~~~
\epsfig{file=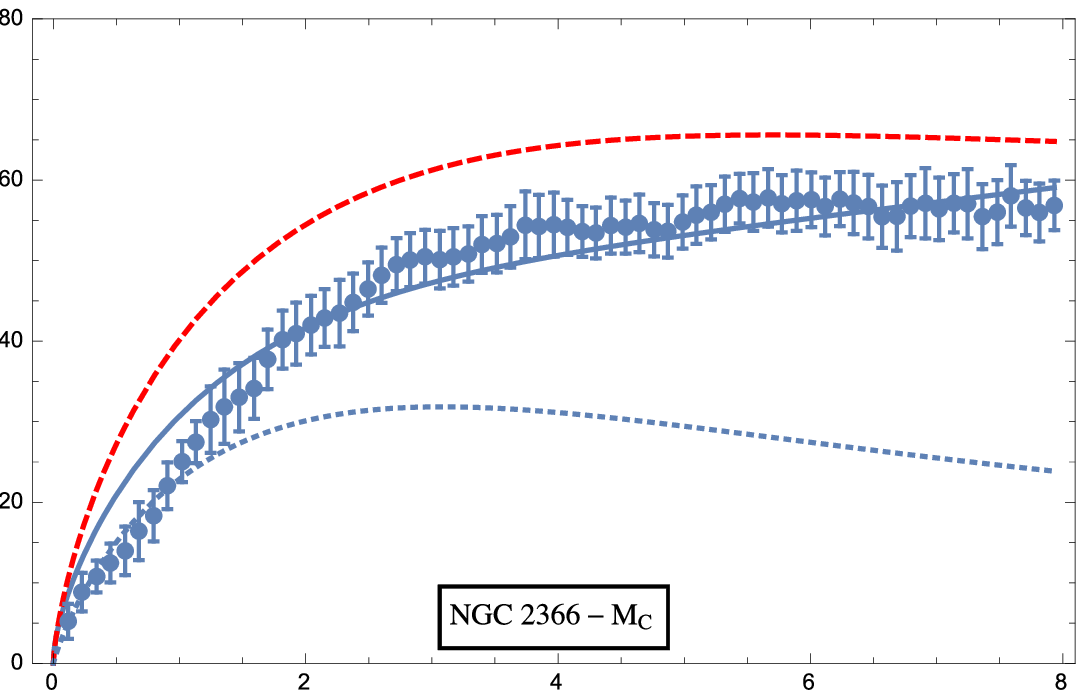,width=1.5825in,height=1.2in}~~~
\epsfig{file=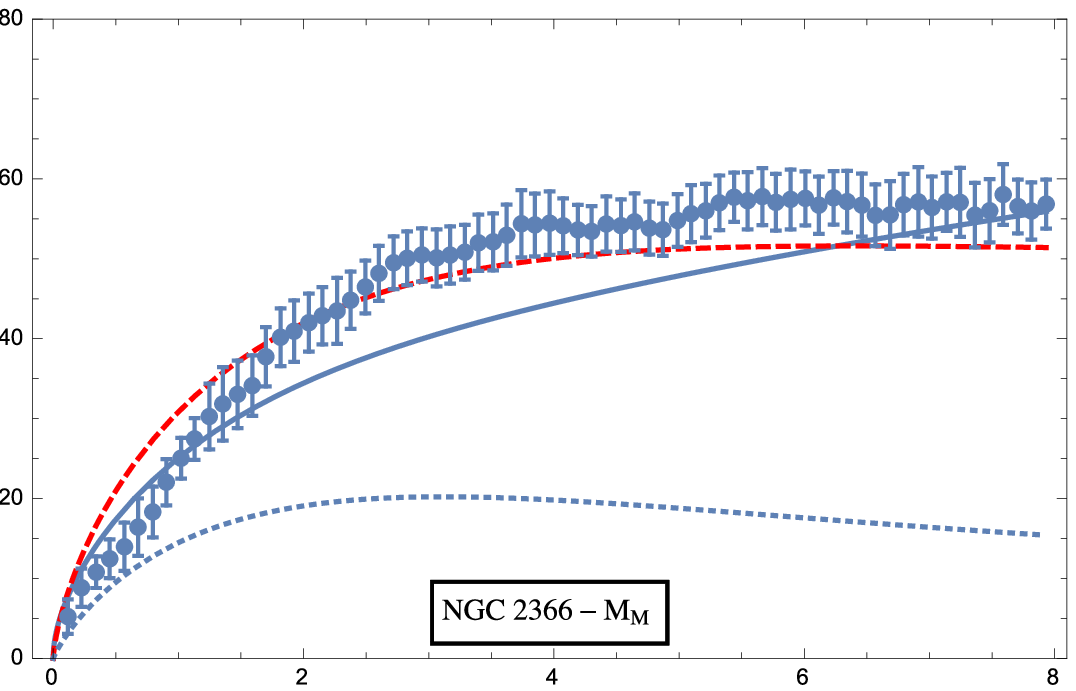,width=1.5825in,height=1.2in}\\
\smallskip
\epsfig{file=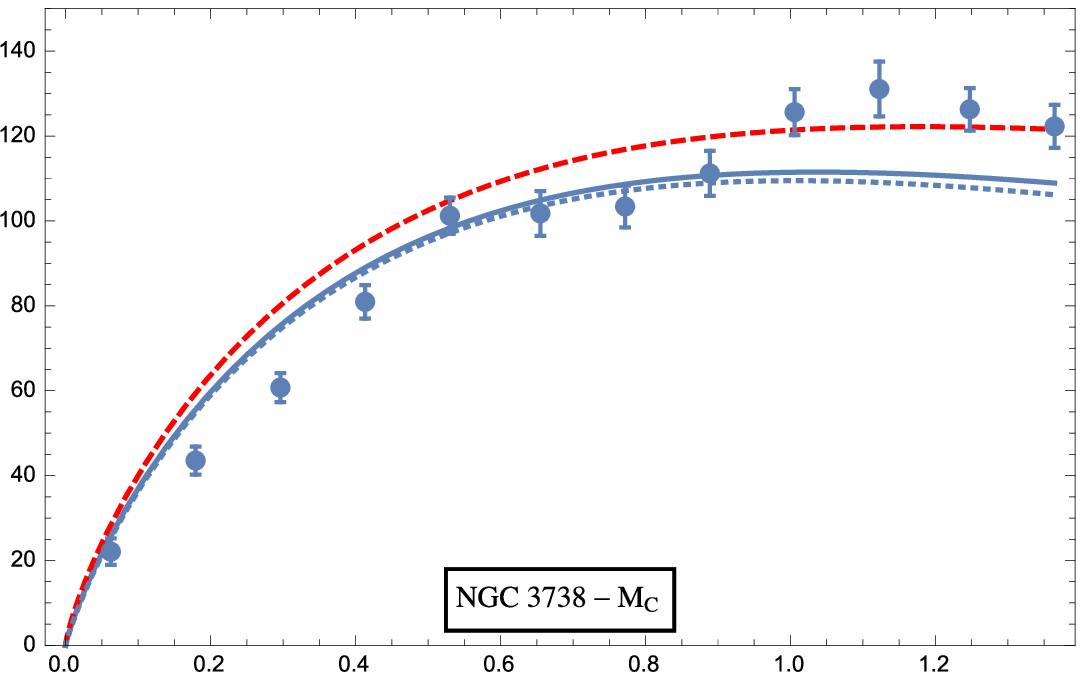,width=1.5825in,height=1.2in}~~~
\epsfig{file=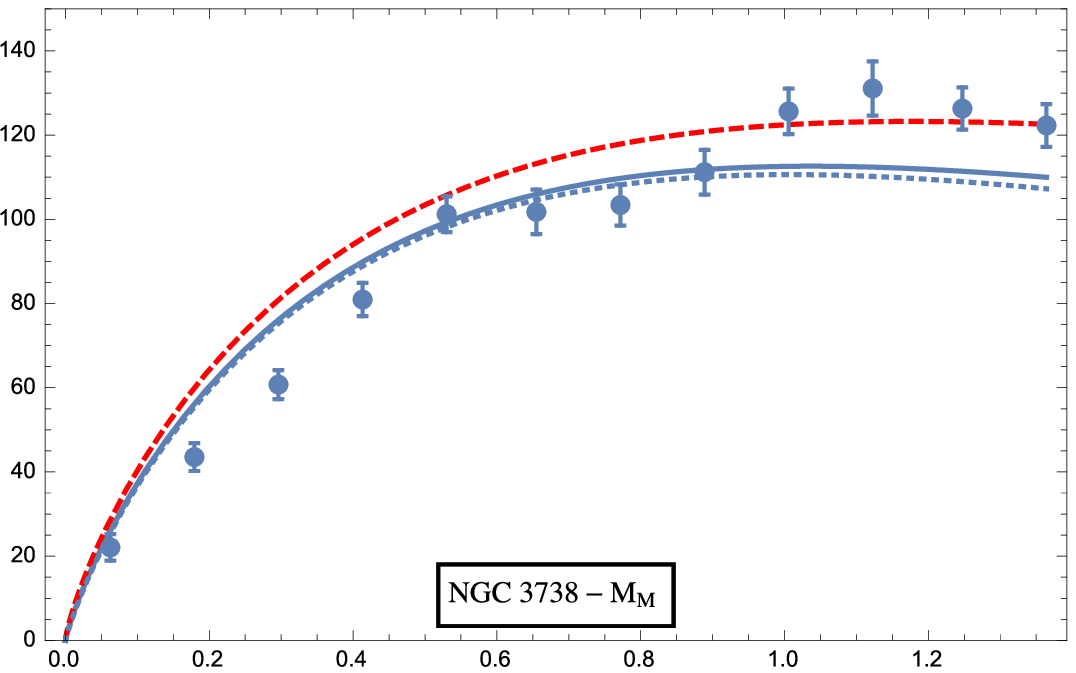,width=1.5825in,height=1.2in}~~~
\epsfig{file=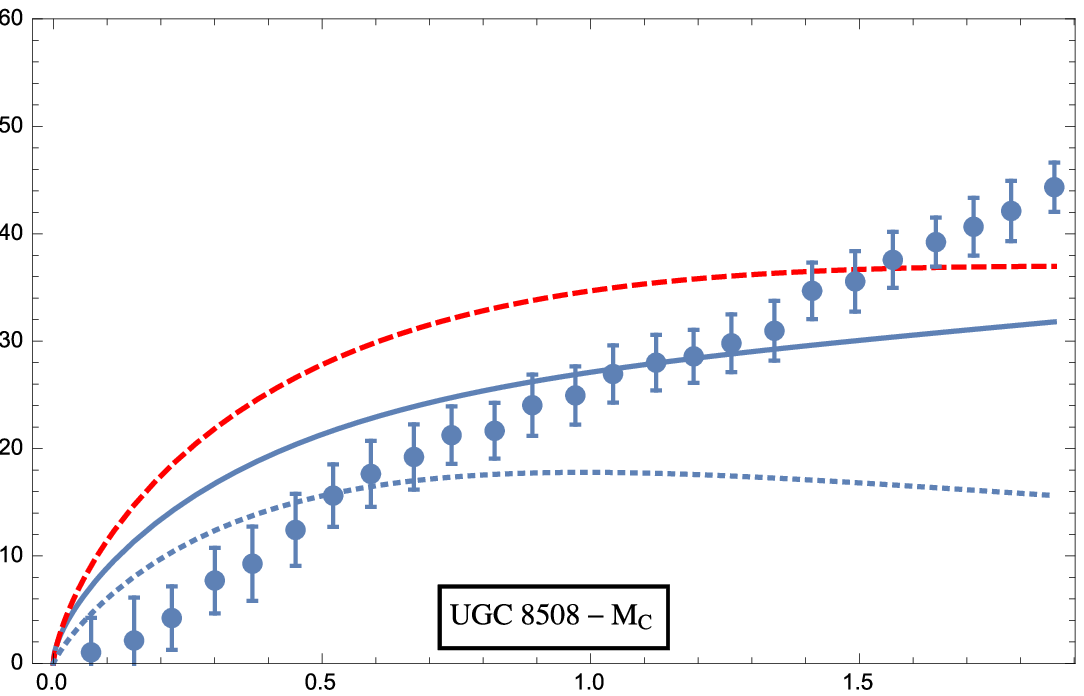,width=1.5825in,height=1.2in}~~~
\epsfig{file=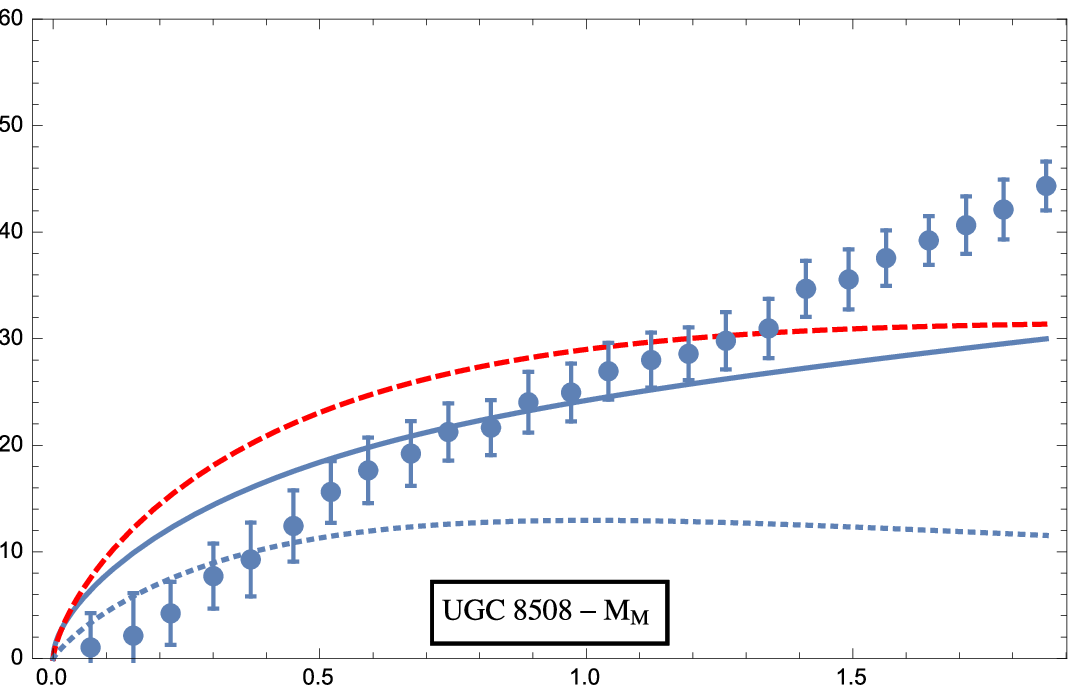,width=1.5825in,height=1.2in}\\
\smallskip
\epsfig{file=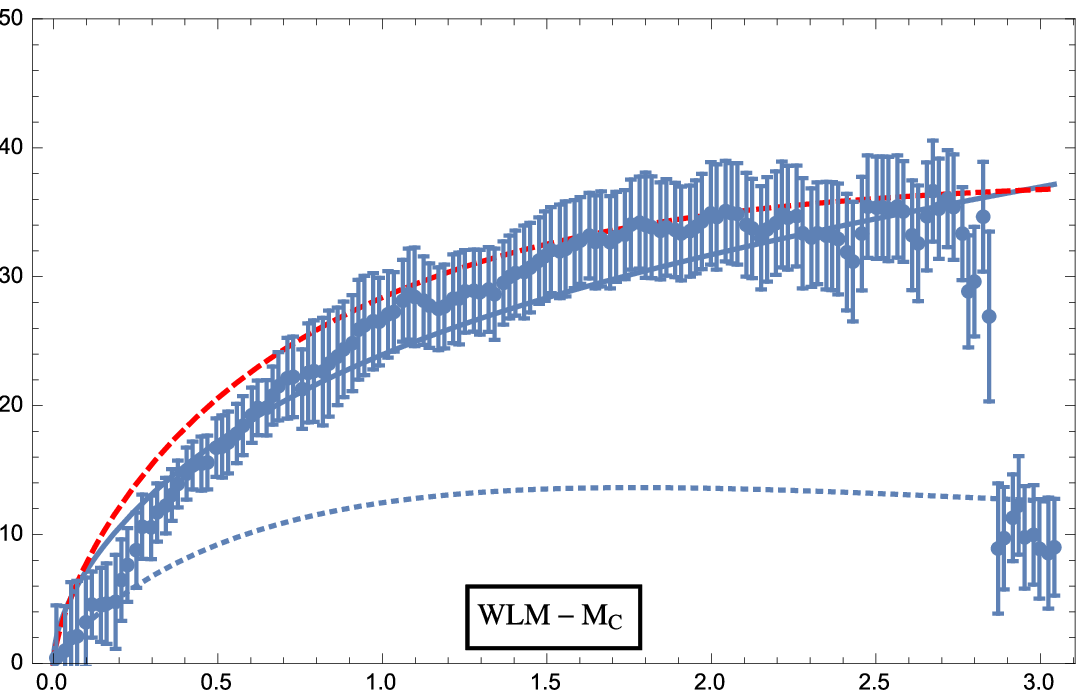,width=1.5825in,height=1.2in}~~~
\epsfig{file=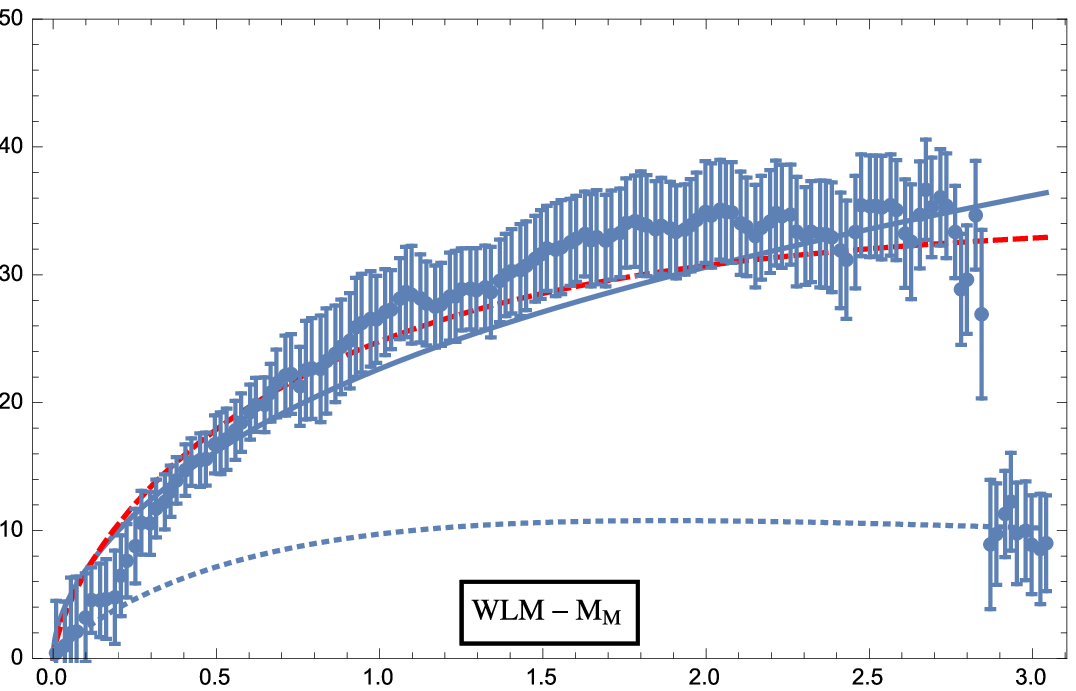,width=1.5825in,height=1.2in}~~~
\epsfig{file=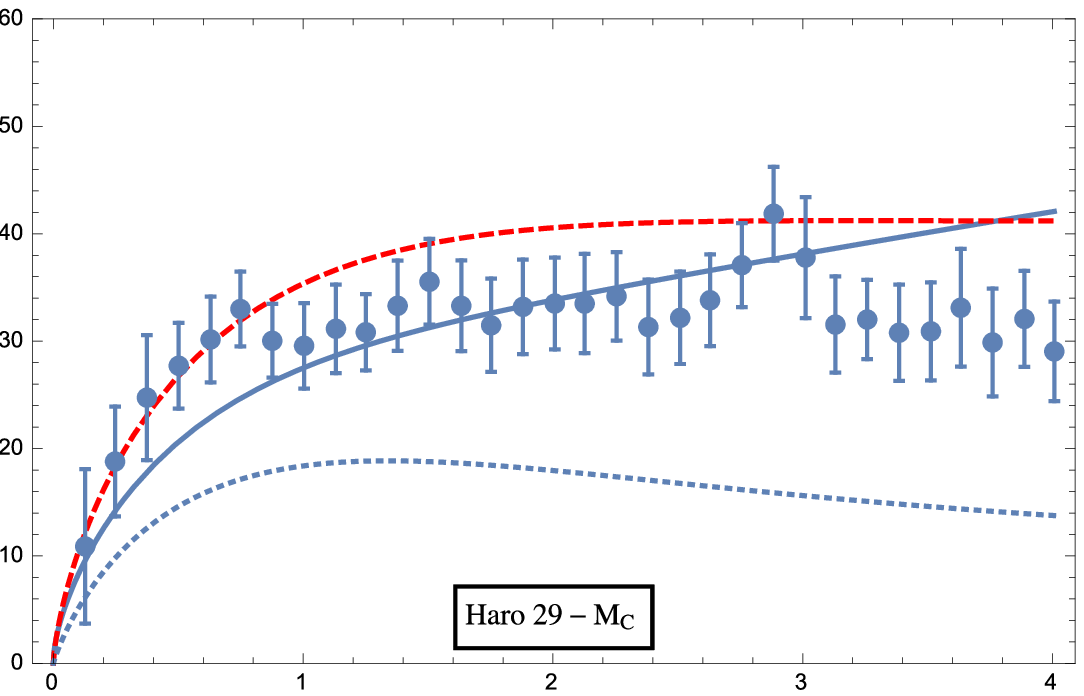,width=1.5825in,height=1.2in}~~~
\epsfig{file=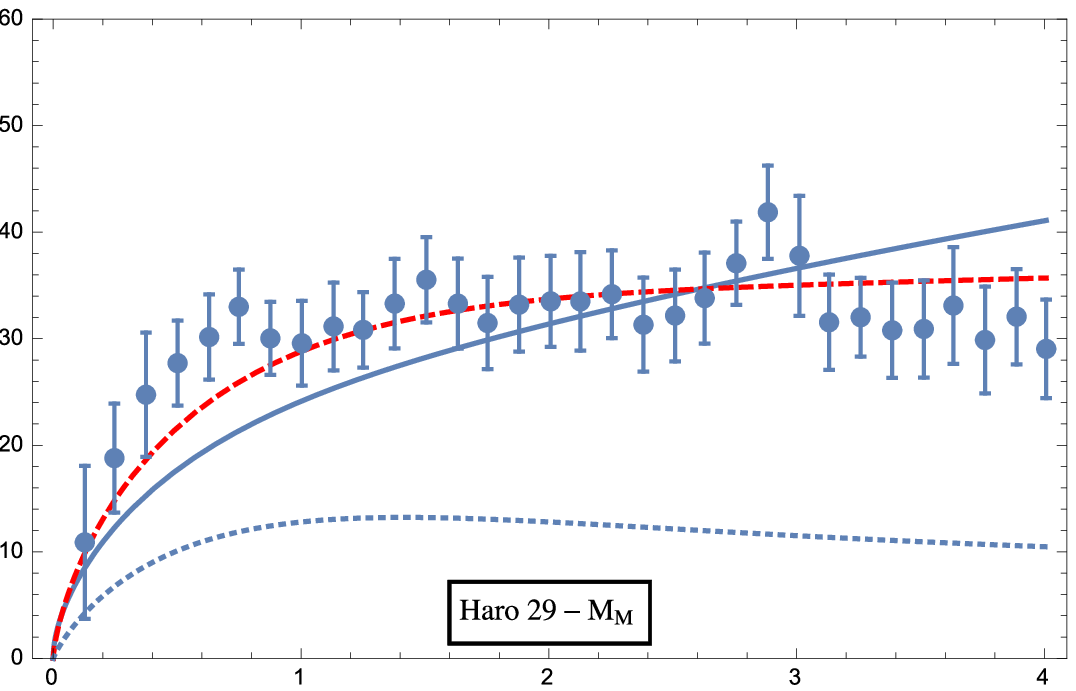,width=1.5825in,height=1.2in}\\
\smallskip
\epsfig{file=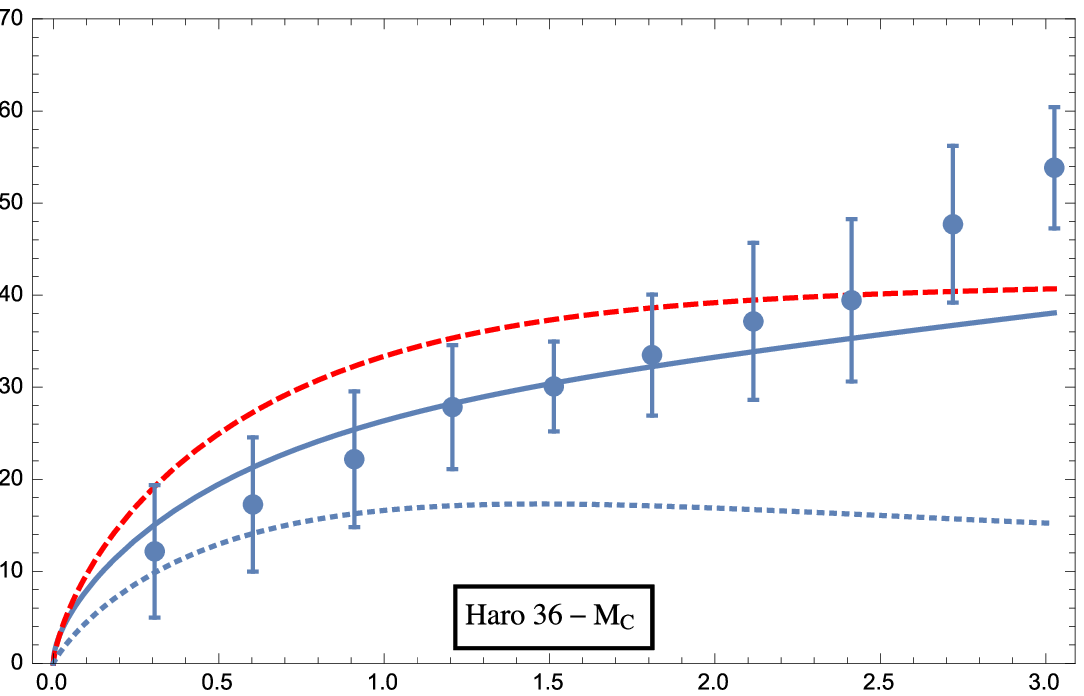,width=1.5825in,height=1.2in}~~~
\epsfig{file=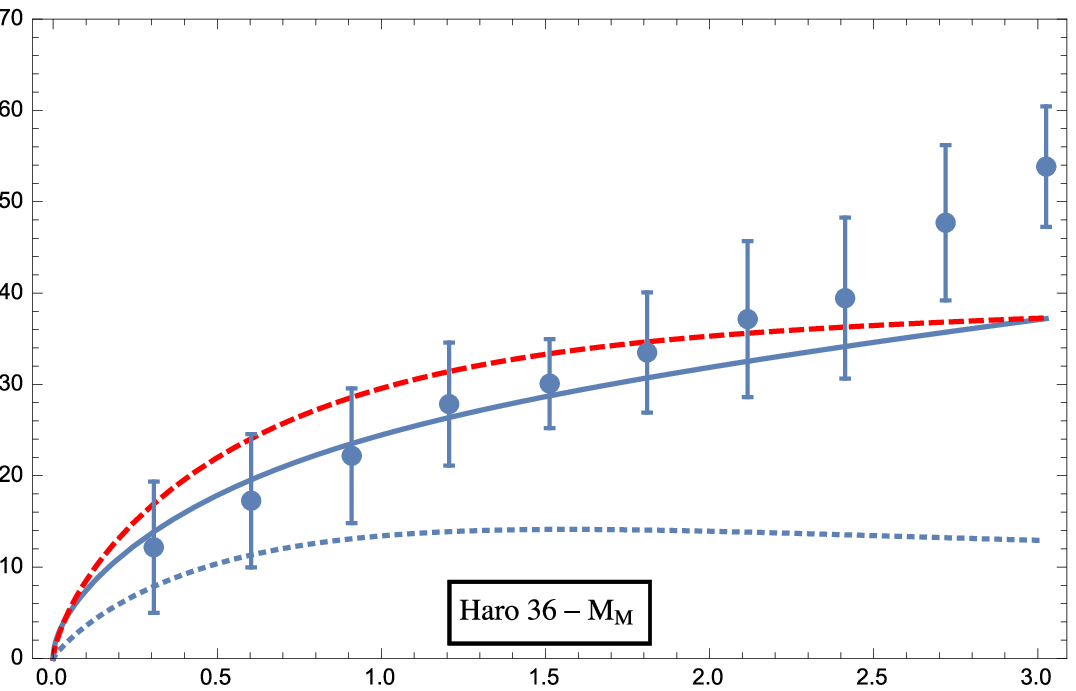,width=1.5825in,height=1.2in}\\
\medskip
\caption{Fitting to the rotational velocities (in ${\rm km}~{\rm sec}^{-1}$) of  the LITTLE THINGS  galaxy sample with their quoted errors as plotted as a function of radial distance (in ${\rm kpc}$). For each galaxy we have exhibited the contribution due to the Newtonian term alone (dashed curve), the full blue curve showing the conformal gravity prediction and the full red curve showing the MOND prediction. No dark matter is assumed.}
\label{rot3}
\end{figure*}

In Figs. \ref{rot1}-\ref{rot3} we present the fits to the data for conformal gravity, MOND and General Relativity.  All of the relevant input parameters are listed in Table \ref{t1} and were obtained from \cite{lTHINGS}, with the exception of the estimated distances.  Since many alternative gravitational models are quite sensitive to distance estimates, one must be careful not to use distance as a free parameter.  Instead, we adopt a uniform method of adopting the estimated distance from the mean of the NASA Extragalactic Database (NED) so as to not provide any bias.  In Figs.  \ref{rot1}-\ref{rot3}, we provide two fits for each galaxy in the LITTLE THINGS sample.\\

In this work, we fit each galaxy for the best fit mass parameter for each theory, and show how both theories perform with the output masses.  For example, in Fig.  \ref{rot1} the nameplate of the first galaxy reads {$CVN-M_C$}, and the neighboring fit reads {$CVN-M_M$}.  These represent the rotation curves for the galaxy CVN, with conformal gravity's preferred mass $(M_C)$ followed by MOND's preferred mass  $(M_M)$ respectively.  In these fits, no other input parameters were changed in the side by side comparison, which allows us to view both theories predictions in a comparative way. In Table \ref{t1} , we report the respective mass to light ratios (with the same subscripts) for the modeled masses.  In the rotation curve fits, we see that the alternative theories can capture the rotational velocity without the need for dark matter, while still providing acceptable mass to light ratios for low surface brightness dwarfs.  Since both theories are using an analytic formula, exotic point by point behavior cannot be captured by eqs. (\ref{total}) and (\ref{MondInt2}).  However, it is important to stress that since the missing mass problem is typically seen outside the inner regions, it is the predictions around the last data points which are most important for this study.  Further, it should be noted that some of these dwarfs are part of companion systems, which can lead to tidal effects, thus making it difficult to provide a reliable stand alone fit without outside influence.  Since this work is meant to show how MOND and CG model a specific self contained set of published data, the authors chose not to eliminate any galaxies from the sample, and instead show the fits to the data as published.  Individual analysis of some of these dwarf galaxies will be addressed in a future work. \\

The LITTLE THINGS survey consists of dwarf galaxies, some of which are irregulars with very high stellar formation rates.  These irregularities, as shown in \cite{lTHINGS}, lead to many interesting features when deriving a rotation curve.  Many of the LITTLE THINGS dwarfs are not only gas rich, but completely gas dominated.  This leads to the gas being almost entirely responsible for the General Relativity rotation contribution, and the galactic disk having little affect on the overall fit.  Typically these gas dominated galaxies also share the trait that the overall rotational velocity is quite low across the measured radii.  Such  galaxies can provide insight to some of the similarities and differences between CG and MOND, and will be discussed below in Section \ref{diffs}.  For this sample, we identify the following eight galaxies to fit the gas dominated and slow rotating criteria:  CVN, DDO 43, DDO 53, DDO 126, DDO 133, DDO 154, DDO 210 and DDO 216.  In this work we show the fits to these eight galaxies (along with the rest of the sample) in Fig.  \ref{rot1}-\ref{rot3}, with the relevant input parameters in Table \ref{t1} .   These galaxies will be denoted with an (*) in Table \ref{t1}  for identification purposes.   Aside from these criteria, the authors in \cite{lTHINGS} identify difficulties in deriving some of the rotation curves of particular galaxies, leading to lower quality factors.  A selection of these factors from the original sources is summarized below.
\subsection{Selected Galaxy Information} \label{sel}
When measuring rotation curve data, there are many systematic issues that affect the accuracy of the information.  In \cite{lTHINGS}, the authors gave a comprehensive discussion of each of the galaxies in the entire sample.  Although both CG and MOND capture the essence of the rotation curves across the sample well, there are occurrences where it is useful to have the relevant information about the galaxy to understand why a fit may not be as precise.  We have compiled some of the relevant information below for convenience.\\

CVN (UGCA 292) is a gas rich dwarf galaxy in the Canes Venatici constellation.  This galaxy has been shown to have a  high gas mass fraction and low metallicity suggesting that it is highly un-evolved with an extremely low surface brightness (\cite{UGC292}).  From a fitting point of view, this galaxy is unique with its extremely low rotational measured velocities. The models shown however, provide a sufficient fit despite this property.\\

DDO 133 (UGC 7698) is a large, faint galaxy in the Canes Venatici cloud whose extremely low surface brightness shows high stellar evolution \cite{UGC7698}. Due to its lack of tidal motions, this galaxy is assumed to be an isolated system which is typically beneficial for obtaining a rotation curve.  However, this rotation curve was studied by one of us in the past, and can be shown to have a lower than quoted inclination (\cite{obriendwarfs}).  Hence, we use the inclination that was previously studied and published to alleviate any discrepancies. \\

DDO 154 (UGC 8024) is an asymmetrical dwarf galaxy and is a member of the Coma Benenices star constellation. It has been proposed that it is a physical companion of the M64 group. The galaxy is very dim, presenting a low star formation rate and contains a very large amount of atomic hydrogen (\cite{UGC8024}). The galaxy is also hypothesized to have a large dark matter to baryon ratio and hence is extremely important to this work.   This galaxy has been studied in CG (\cite{fitting}) after the initial THINGS survey was completed  by \cite{Walter2008}.  It can be seen in this most recent survey that, the alternative gravity models can circumvent the need for the quoted high dark matter to baryon ratio.   \\

DDO168  (UGC 8320) is a galaxy in the Canes Venatici group and is a probable pair with UGC 8308 and sometimes included in the M51 group (\cite{UGC8320}). The galaxy has a large number of dispersed HII regions, resulting in an unusual shape to the rotation curve.  The essence of the rotation curve is captured in the fits, but due to the overall structure, one cannot expect a high resolution fit.\\

DDO 210 is located in the Aquarius star constellation. It was one of the original galaxies discovered by \cite{vdb} of the David Dunlap Observatory catalogue (DDO) survey. This aquarius dwarf is extremely faint, and has a very low surface brightness, though it contains several varying HII regions. The surface brightness is so low, that it is not possible to determine or map any spiral arms. There is inconsistency in the photometry data because of a strong foreground contamination and the presence of several bright foreground stars makes it difficult to derive a reliable rotation curve (\cite{AqD}, \cite{AqDpro}).  With these comments, we include the galaxy in our fits, but we do not make attempts to resolve the data with the predictions of the alternative models.\\

IC 1613 or UGC 0668 is a Local Group galaxy in the Cetus constellation and has a morphological classification of IBm (Irregular Barred-Magellanic). It exhibits low surface brightness and contains very little dust making it relatively transparent, excluding foreground or background contamination (\cite{UGC0668}).  The presence of the bar and transparency make this galaxy extremely difficult for a reliable fit.\\

\subsection{Differences in the CG and MOND Mass Models} \label{diffs}
Although both CG and MOND are alternative gravitational theories that explain rotation curve physics without dark matter, the mechanisms for circumventing missing matter is quite different in each theory.    MOND requires a centripetal acceleration in the form of $a_{0} = 1.35 \times 10^{-8} ~cm\,s^{-2}$ for dwarf galaxies.  In squaring eq. (\ref{total}) and neglecting the quadratic term (which only contributes significantly in large spirals), we see CG also yields an acceleration, with both a $\gamma^*$ and $\gamma_0$ term, with the $\frac{\gamma_0 c^2}{2}=1.4*10^{-9} cm\,s^{-2}$ being the universal term, which we note is not terribly far off from the MOND value.  However, since the $\gamma^*$ term is still coupled to mass (both disk and gas), this leads to departures from universal motion and is instead tied to the baryonic content of the particular galaxy. 
Despite both theories possessing a universal acceleration scale,  the presence of the ($\gamma^*$) term in CG requires us to look at some specific galaxies to highlight the behavior of the two theories.  \\
As a first illustration, we compare the two fits of CVN in Fig.  \ref{rot1}.  Since the preferred masses (as listed in Table \ref{t1} ) are determined using a $\chi^2$ minimization procedure, the preferred CG mass comes out to be about 10 times less than the preferred MOND mass.  However, this is a gas dominant galaxy, where the gas mass (unchanged in the two fits) is about the same as the mass found using MOND.  Hence, when fitting, this serves to bump the newtonian contribution to essentially twice the gas contribution.   Using this mass generates the MOND preferred fit for CVN, where we can see that in eq.  (\ref{MondEq2}) at small distances, $V_{MON}(R)\approx \sqrt{2}V_{NEW}$.  Hence for the MOND fit of CVN, we see that at around $r=.2kpc$, the disk and gas rotation contribution alone yields $V_{MON}\approx2 km\,s^{-1}$, generating the predicted value of the MOND fit (red curve).  However, this same fit shows how the CG theory reacts at small distances with small masses.  In comparison with the CG predicted mass (smaller by a factor of ten), there is little difference in the CG fit (blue curve) across both panels.  This is because at such a small distance with small mass, the linear $\gamma_0$  term alone has already begun to dominate, such that if we neglect the mass altogether, then eq. (\ref{total}) becomes $v_{cg}\approx c \sqrt{\frac{\gamma_0}{2}}$.  At the same distance of $r=.2kpc$, this yields $v_{cg}\approx 9.23 km\,s^{-1}$.  Hence, the non baryonic term is already completely dominant.  It should be noted that since changing the mass by a factor of ten provides little change for the CG fit, the authors have kept the smaller mass for the CG fit  to preserve consistency while using the minimum $\chi^2$ predicted mass.  Similar features can be found in the very next galaxy in the sample, DDO 43. In DDO 43, the non baryonic term ($\gamma_0$)  is once again dominant in the inner region, and hence changing the mass by a small factor does not change the fit dramatically. Since the $\gamma_0$ term in eq. (\ref{total}) is galaxy independent,  the slow rotation coupled with a small termination radii (such as CVN) causes the CG formula to generate a small disk mass, whereas MOND will predict a multiple of the gas mass.\\
To contrast the analysis of CVN, we can use DDO 154 as another case study.  In this galaxy, there is a small velocity in the inner region, however the velocity rises slowly to over 5 times that of the inner region, and has a large termination point compared to CVN.  In Table  \ref{t1}, we see that the masses predicted by MOND and CG are quite similar, resulting in similar fits across the two panels.  Hence, despite the dominance of the inner region by the $\gamma_0$ term in CG, there needs to be a non trivial newtonian contribution due to the size of the galaxy, and both theories show very similar overall curves for the two chosen masses.  The two theories begin to deviate as expected since CG has the extra luminous term ($\gamma^*$) for both the gas and disk.\\
As a last illustration, we turn to DDO 168.  As seen in Table  \ref{t1}, the two predicted masses are of the same order of magnitude, but off by a factor of 2.  This galaxy, like DDO 154 has larger overall velocities and terminates at a distance a few multiples of the optical scale length. However, in this galaxy, the disk in both fits dominates the gas.  In this galaxy, CG predicts a lower overall disk mass (with an unchanged gas mass) due to the extra luminous contribution of the $\gamma^*$ term.  In summary, since the input parameters of a given galaxy are unchanged across the two fits, the size and observed velocities of the given galaxy dictate how the mass will be predicted.  It should be noted however, that if there is uncertainty in some of the observational measurements (such as inclination), this would lead to a different observed velocity being reported, and hence a different mass being predicted in the modeling described (see for example \cite{obriendwarfs}).

\subsection{Tully Fisher relationship}

\begin{figure}
    \subfigure[]{%
 \epsfig{file=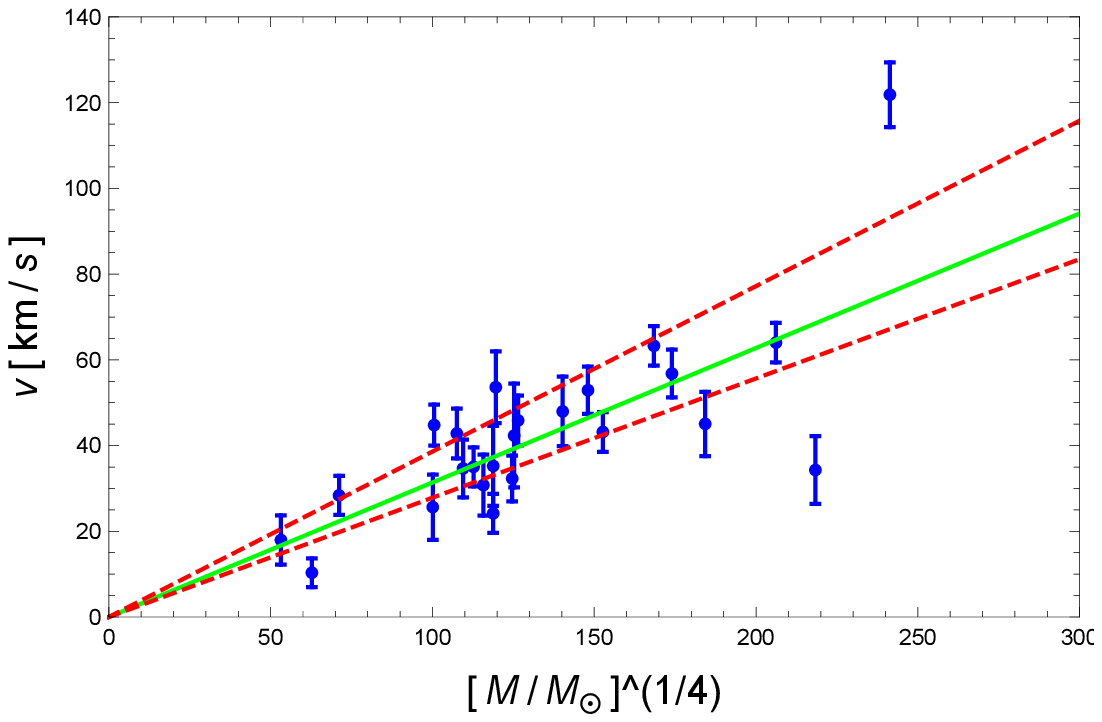,width=3.2in,height=2.05in}
  \label{tfcg}}
   \subfigure[]{%
 \epsfig{file=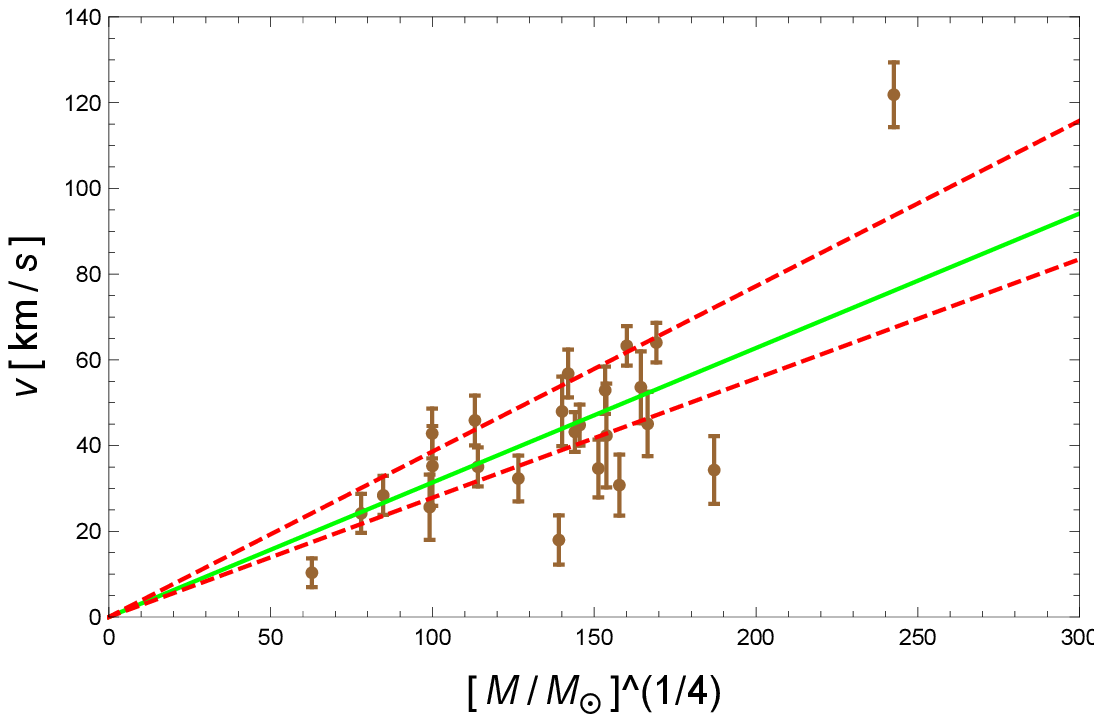,width=3.2in,height=2.05in}
  \label{tfmond}}
  \caption{ Fig. \ref{tfcg} shows the Baryonic Tully Fisher plot for CG while \ref{tfmond} shows the same plot for MOND, where in both figures the last observed velocity was used.}
   \label{tf2}
\end{figure}

The well known Tully Fisher relation \cite{Tully1997a}, later revisited by \cite{bftully} viz.,
\begin{equation}
 v_{OBS}^4 \propto  M_{tot},
 \label{tf}
 \end{equation}
is an empirical relationship where the assumed luminous mass {$M_{tot}=M_d+M_{gas}$}  is shown to be proportional to the observed velocity to the fourth power.  Eq. (\ref{tf}) is typically applied in  the flat region of a rotation curve or the terminal point in dwarf galaxies. In Fig. \ref{tf2} we present the plots of how the mass prediction of each respective theory as given in Table \ref{t1} , relates to the last observed velocity data point for each galaxy in LITTLE THINGS. In order to highlight the Tully Fisher relationship with the velocities' natural error, we rescale the plot in Fig. \ref{tf2} to show $v \propto M^{\frac{1}{4}}$.  To show how these plots relate to previous work on the baryonic Tully Fisher relationship, we can write the proportionality in terms of a constant as $v=AM^{\frac{1}{4}}$.  In Fig. \ref{tf2}, we show the last data points as fit to the exact $v^4$ relationship with their respective errors for both CG and MOND (Fig. \ref{tfcg} and Fig. \ref{tfmond} respectively).  The green line in these fits represents the the best fit value for the constant A across the data points.  The red dashed plots show a range of values for the constant A, which correspond to a range of values found in \cite{bftully}.
We see that the assumed masses for the two theories on a point by point basis capture the Tully Fisher plot in Fig. \ref{tf2}, with the aforementioned slow rotating galaxies providing the majority of the outliers in both theories.  It should be noted that dwarf galaxies as a whole tend to be problematic overall for the Tully Fisher relation, which was the impetus for the inclusion of gas mass into the empirical relation.  In \cite{bftully}, the authors experiment with values of the exponent in eq. (\ref{tf}) ranging from 3.0-4.5.  In this work, we restricted the exponent to the exact value of 4 to show that we can get agreement within the range of A values across the LITTLE THINGS sample. \footnote{In \cite{bftully} the authors explore the Baryonic Tully Fisher relation given as $M=Bv^x$.  When restricting $x=4$, the authors arrive at a range of values of $50   \frac{M_\odot}{(km/s)^4}\leq B \leq 150\frac{M_\odot}{(km/s)^4}$.  These B values correspond to the bounds shown in the red dashed plots in Fig. \ref{tf2} when renormalized to $v=AM^{\frac{1}{4}}$.} In Fig. \ref{tfcg}, we note the four outliers as DDO 210, DDO 46, DDO 168 as the galaxies below the accepted value range, and NGC 3738 as the one way above the range, and for Fig. \ref{tfmond} we have the same outliers with the addition of IC 1613.  In the analysis above in Section \ref{diffs}, DDO 46 and DDO 168 were discussed and we note are two of the galaxies classified as gas dominant.   DDO 210 is one of the galaxies highlighted by Oh et al. as being quite difficult to isolate and fit and has one of the smallest and slowest rotation curves observed.  Lastly, IC 1613 was noted in Section \ref{sel} by Oh et al. as to contain a bar, as well as having interference from the background and foreground due to its transparent nature, making it difficult to produce a reliable fit.  Further, Table \ref{t1} shows that the predicted masses for IC 1613 by a factor of about twelve (see Section \ref{diffs} as to why the masses can vary drastically), which highlights why it is only an outlier in the MOND fit and not CG.  Since  the Tully Fisher relation relates mass to velocity, if we have difficulty establishing a good velocity, the mass model will suffer as well.  Within this survey, NGC 3738 presents itself as a galaxy with a very small rotation curve (radially), but also as a galaxy that rises to a high rotational speed quite quickly.  In this galaxy, the Newtonian contribution dominates over the additional contributions of MOND and CG respectively.  However, if there is a discrepancy in the distance to the galaxy (The NED shows $1.8 Mpc \leq D \leq 5.3 Mpc$), this could lead to a variation in the mass modeling resulting in a higher mass prediction and thus falling off of a Tully Fisher plot.  In the NED listing, there are various methods in the literature for distance estimation, including a distance derivation based on the assumption of the target galaxy obeying Tully Fisher.  Since we have chosen to use NED averages for consistency in this work, and not favor any particular distance for a given galaxy, we would expect that galaxies could fall outside of our predicted range in the Tully Fisher plot.  In a future work, the authors will highlight particular galaxies for which Tully Fisher is assumed to generate a distance estimate versus those whose distances are measured by other means (such as cepheids) in order to see the affects on mass modeling and the relation to the Tully Fisher relation.

\subsection{Velocity Dispersion Counts}
\begin{figure}
\epsfig{file=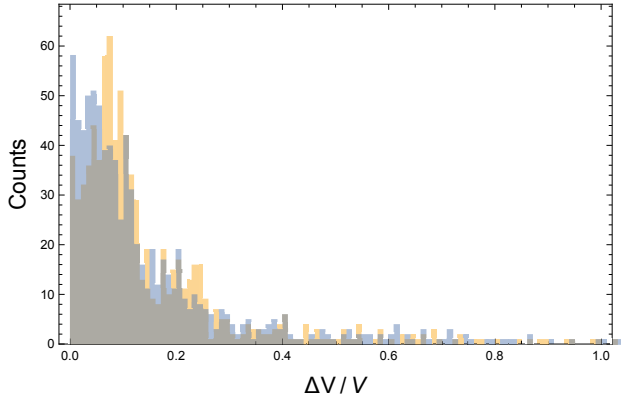,width=3.2in,height=2.05in}\\
\smallskip
\caption{The velocity dispersion counts for the entire sample on a point by point basis.  The blue bars show the counts for CG and the yellow bars show the counts for MOND.}
\label{histo}
\end{figure}

The rotation curves shown in Figs.  \ref{rot1}-\ref{rot3} provide adequate fits to the data for the given masses, but it is difficult to ascribe a preferred theory in some cases.  To provide an overall summary of how the two models behave across the entire sample, Fig. \ref{histo} shows a histogram of the velocity discrepancy {$\Delta V/V_{obs}$} for CG and MOND models respectively on a point by point basis.  
Over the entire sample, we see that both theories provide a good agreement to the data,  and show comparable average counts across the sample in velocity dispersion. \\

\section{Centripetal Accelerations}

In a recent article, \cite{mcgaugh2016}, used rotation curve data to highlight universal centripetal accelerations present in spiral galaxies.  A major conclusion of their work was the possibility of new dynamical laws as the cause of universality as opposed to dark matter.   Thus, we can use the LITTLE THINGS sample as a further test of this possibility through the lens of conformal gravity and MOND.\\

\begin{figure}
    \subfigure[]{%
 \epsfig{file=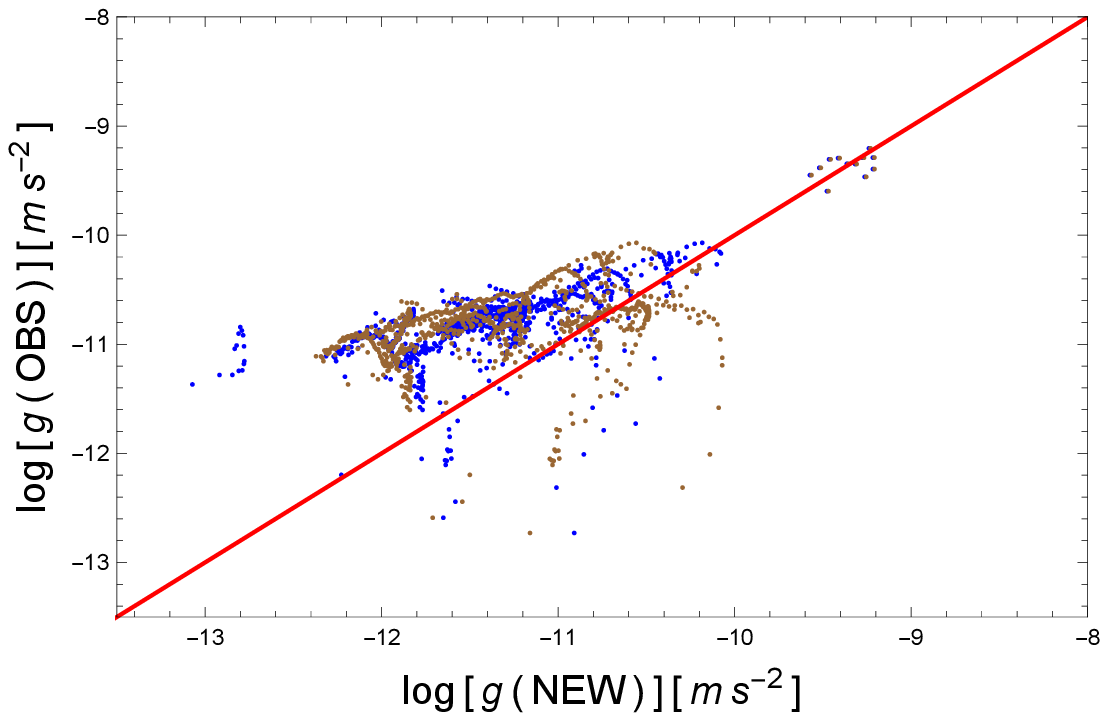,width=3.2in,height=1.85in}
  \label{gbar}}
   \subfigure[]{%
 \epsfig{file=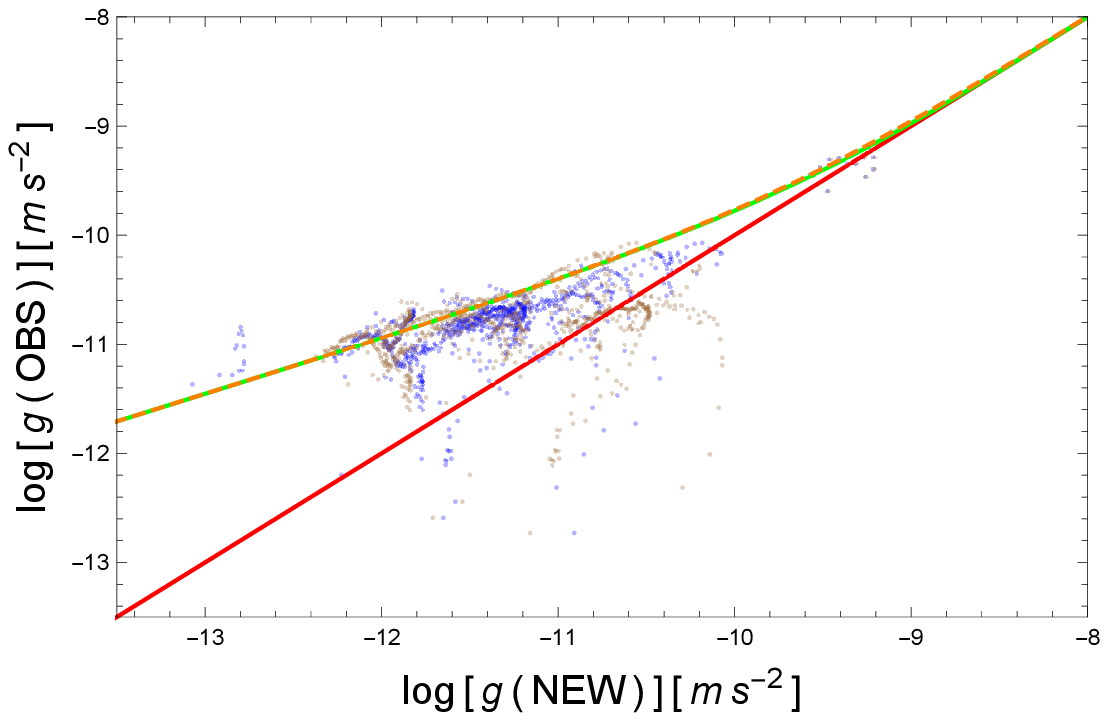,width=3.2in,height=1.85in}
  \label{mondfig}}
    \subfigure[]{%
 \epsfig{file=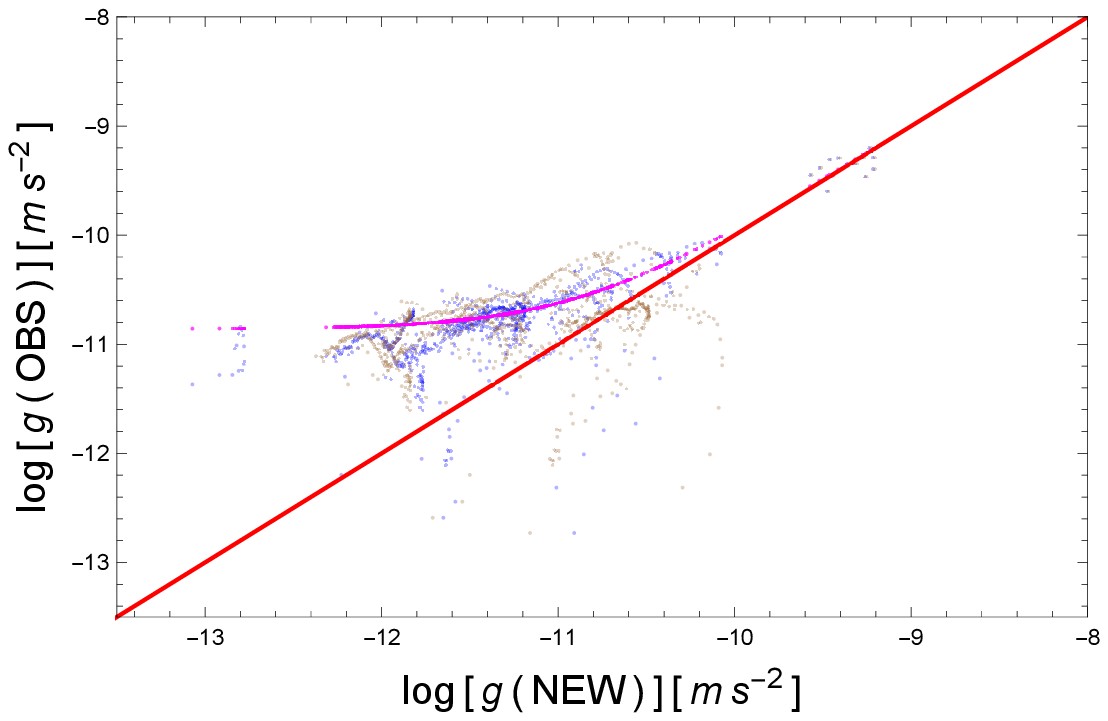,width=3.2in,height=1.85in}
  \label{gcg}}
 \caption{Fig. \ref{gbar} shows the point by point comparison of $(g_{NEW},g_{OBS})$, with the blue points using the mass predicted by CG and the brown points using the mass predicted by MOND.  Fig. \ref{mondfig} shows the same points with $g_{MON}$ shown in the dashed orange line and $g_{MLS}$ shown in the green line. Fig. \ref{gcg} shows Fig. \ref{gbar} with the CG prediction overlaid in purple. The line of unity is shown in red in all three plots.}
   \label{gobs}
\end{figure}

Universality in rotation curves has already been noted by \cite{Persic}.  As seen in the last column of Table  \ref{t1}, the centripetal acceleration of these points (scaled to $\frac{1}{c^2}$), trends to a value within an order of magnitude around unity.  This feature has been noted in \cite{fitting}, and is important outside of the LITTLE THINGS survey as the trend seems to transcend morphology for the full set of 150 galaxies studied by CG.  However, as in \cite{mcgaugh2016}, we can extend this trend to the entire point set of the LITTLE THINGS survey.  To this end, we present in Fig. \ref{gbar} the observed centripetal accelerations $g(OBS)=\frac{v^2_{OBS}}{R}$ versus the predicted values of the baryonic matter alone, $g(NEW)=\frac{v^2_{NEW}}{R}$, with $v_{NEW}$ given in eq. (\ref{gr}), where $v_{OBS}$ and R are the observed central velocity at the given distance from the galactic center respectively.  We note that since g(NEW) is mass dependent, we plot both g(NEW) with the mass values predicted by CG and MOND in blue and brown respectively over the entire 850 points of the LITTLE THINGS sample.  Since these are dwarf, Low Surface Brightness (LSB) galaxies, these points only populate a small regime of that noted in  \cite{mcgaugh2016}.  To see how the two theories in question accommodate centripetal accelerations, we compare the predictions in each.  For MOND, we can plot the prediction in eq. (\ref{MondEq2}), 
\begin{equation}
g(MON)=g(NEW)+g(NEW)\left(\frac{\sqrt{1+\frac{4a_0}{g(NEW)}}-1}{2}\right)\\
\label{gmon}
\end{equation}
where we use the g(NEW) as predicted by the MOND mass given in Table \ref{t1} .  This equation must admit a similar result to the function fit in \cite{mcgaugh2016}, viz,
\begin{equation}
 g(MLS)=\frac{g(NEW)}{\left(1-\exp(\frac{-(g(NEW)}{g_\dagger})^{\frac{1}{2}})\right)},\\
 \label{gmls}
 \end{equation}
 due to the asymptotic nature of both eq. (\ref{gmls}) and eq. (\ref{gmon})  and the fact that $g_\dagger=a_0$.  The fit to the data for these two functions are shown in Fig. \ref{mondfig}.
However, in CG, the new physics described by McGaugh is captured by the combination of both local and global effects of gravity.  We see that although Fig. \ref{gcg} predicts a narrow band, it captures the average of the width of the CG data (blue points) on this scale, whereas in Fig. \ref{mondfig}, the relevant MOND curves capture the MOND masses well (brown points). Each theory however behaves differently in the regime of small g(NEW). In Fig \ref{mondfig} and Fig. \ref{gcg} respectively, we see that the MOND curves fall slightly below the data, whereas due to the asymptote of CG at accelerations no smaller than $10^{-11}$, the CG fit is slightly above the data. Since the plots shown here (to mirror those shown in \cite{mcgaugh2016}) do not include error bars on the velocity points, we can  conclude that the two theories both can accommodate the universal centripetal accelerations with neither being preferred.  We note that both theories predict a narrow band over the entire acceleration regime, and the significance of the width of the band is discussed in \cite{obrientom}.  In order to best distinguish between the two theories, one would have to include galaxies outside this small sample and include large spirals instead of just dwarfs to see the behavior in the entire acceleration window.  Preliminary work on this larger analysis has already been started and the bright spirals form a smaller, narrower band, and will be discussed in a future work.

\section{Conclusion}

Alternative theories once again are shown to predict galactic rotation curves without the need for dark matter. Further, due to the recent activity in the literature about universal centripetal acceleration, we provide insight into how these two theories fit the observed data.   The LITTLE THINGS survey adds a robust set of dwarf galaxies to the increasingly large population of galaxies that alternative models of gravity have been able to successfully fit.  This paper captures how two of these theories make predictions for a given galaxy with the same set of input parameters in a single fit, and do so without dark matter.  The LITTLE THINGS sample alone is not enough to rule out one theory in favor of the other, but the work contained here has shown how MOND and CG handle dwarf galaxies, and can be extended to a more diverse set of morphologies in future work.\\
\smallskip

\section*{Acknowledgements}
 J. G. O'Brien and T. L. Chiarelli would like to thank Dr. P. D. Mannheim for his continued efforts in the advancement of conformal gravity. R. J. Moss would like to thank P. McGee, David Miller, and A. Clement for their contribution to \cite{mossobrien} which made the analysis of this work seamless.  J. G. O'Brien and J. Dentico would like to thank Dr. J. Brownstein for his input and insight into MOND.  J. G. O'Brien would like to thank Dr. G. Sirokman and Dr. Douglas Goodman for their helpful comments on this paper. The group would like to thank Dr. S. Oh for sharing the data of the LITTLE THINGS Survey.  This research has made use of the NASA/IPAC Extragalactic Database (NED) which is operated by the Jet Propulsion Laboratory, California Institute of Technology, under contract with the National Aeronautics and Space Administration.

\bibliographystyle{aasjournal}
\bibliography{citations}
\begin{table*}
\caption{Properties of the LITTLE THINGS Galaxy Sample.}
\centering
\begin{tabular}{l c c c c c c c c c c c } 
\hline\hline
Galaxy &$ i$        & Dis.    & Lum. & $R_0$ & $R_l$  & $M_{gas}$   	    & $(M_{d})_C$        & $(M/L)_{C}$ & $(M_{d})_M $ &$(M/L)_{M}$  &  $(v^2 / c^2 R_{l})_{last}$ 	 \\
\phantom{0000000000}   & $(o)$  & (Mpc) & $(10^8L\odot)$ 	& (kpc) & (kpc)  & $(10^7M_\odot)$ & $(10^8M_\odot)$ & $(M_\odot L_\odot^{-1})$	 & $(10^8M_\odot)$ & $(M_\odot L_\odot^{-1})$  & $(10^{-30}cm^{-1})$ \\
\hline 
CVN*	 	  & 66.5  & \phantom{0}3.18   & $0.11$                 & 1.1 & 2.5 & $\phantom{0}2.26$  & $\phantom{0}0.03$ 	& 0.27 &\phantom{0}0.29 &  2.69      & $1.16$  \\
DDO 43*      & 40.6  & \phantom{0}7.40  & $1.54$ 		 & 1.1 & 4.0 & $20.94$ 	               & $\phantom{0}0.32$ 	& 0.21 & \phantom{0}0.47& 0.31       & $0.94$ 	\\
DDO 46      & 27.9  &\phantom{0}8.78  & $2.45$ 		 & 0.9 & 5.1 & $45.77$ 	               & $18.15$ 	                   & 7.42 & \phantom{0}7.67&3.14        & $0.83$ 	\\
DDO 47      & 45.5  &\phantom{0}6.00  & $3.28$ 		 & 1.1 & 8.9 & $62.31$ 	               & $\phantom{0}1.82$   & 0.55 & \phantom{0}0.33&0.10        & $1.62$ 	\\
DDO 50	  & 49.7  & \phantom{0}2.28  & $3.06$ 		 & 0.5 & 6.6 & $\phantom{0}6.08$  & $\phantom{0}1.95$ 	& 0.64 & \phantom{0}1.03&0.33        & $8.92$ 	\\
DDO 53*      & 27.0  & \phantom{0}3.31  & $0.44$                  & 0.6 & 1.4 & $\phantom{0}6.34 $ & $\phantom{0}0.37$  & 0.87 & \phantom{0}0.33&0.77        & $1.69$ 		\\
DDO 70      & 50.0  & \phantom{0}1.59  & $1.01$ 		 & 0.7 & 2.4 & $\phantom{0}5.67 $ & $\phantom{0}0.77$   & 0.76 & \phantom{0}0.43&0.42        & $2.75$ 		\\
DDO 87      & 55.5  & \phantom{0}7.27  & $1.40$ 		 & 2.3 & 7.0 & $25.96$                     & $\phantom{0}2.21$   & 1.58 & \phantom{0}2.94&2.10         & $1.44$ 		\\
DDO 101    & 51.0  &                      11.25 & $4.85$ 		 & 1.1 & 4.1 & $10.82   $ 	               & $16.98$              	&3.53  & \phantom{0}7.11&1.47         &$3.60$ 		\\
DDO 126*    & 65.0  & \phantom{0}4.14  & $1.01$ 		 & 1.0 & 3.4 & $11.73$ 	               & $\phantom{0}0.44$ 	& 0.43 & \phantom{0}0.52&0.51         & $1.30$ 		\\
DDO 133*    & 43.4  &\phantom{0}4.95  & $2.59$ 		 & 1.6 & 4.9 & $25.72$ 	               & $\phantom{0}2.86$   & 1.10 &  \phantom{0}1.73&0.67        &$1.37$ 	\\
DDO 154*	  & 68.2  &  \phantom{0}3.84 & $0.80 $                 & 0.7 & 8.2 & $37.99$                     &  $\phantom{0}0.07$    &0.08&\phantom{0}0.05&  0.02        &$1.01$                \\
DDO 168    & 46.5  &\phantom{0}4.61  & $3.41$ 		 & 1.2 & 4.3 & $29.75$ 	               & $\phantom{0}8.56$ 	& 2.51 & \phantom{0}4.71&1.38          & $1.70$ 	\\
DDO 210*	  & 66.7  &\phantom{0}0.70  & $0.22$                  & 0.4 & 0.3 & $\phantom{0}1.05$   & $\phantom{0}0.05$ 	& 2.11 & \phantom{0}0.04&2.11          &  $1.28$ 	\\
DDO 216* 	  & 63.7  &\phantom{0}0.76  & $0.21$                  & 0.9 & 0.7 & $\phantom{0}0.20 $ & $\phantom{0}0.06$ 	& 0.23 & \phantom{0}3.72&1.52          & $1.66$ 		\\
F-564 V3    & 56.5  &\phantom{0}8.73  & $0.68$                  & 0.7 & 3.7 & $\phantom{0}4.40$   & $\phantom{0}0.79$ 	& 0.11 & \phantom{0}0.53&0.08          & $0.55$ 	\\
IC 10	  & 47.0  &\phantom{0}0.87  & $6.85$ 		 & 0.4 & 0.6 & $\phantom{0}2.20$  & $\phantom{0}1.77$	& 0.26 & \phantom{0}0.78&0.11          & $7.45$ 	\\
IC 1613      & 48.0  &\phantom{0}0.61  & $7.96$ 		 & 0.7 & 0.6 & $\phantom{0}2.20$  & $\phantom{0}1.77$ 	& 0.22 & \phantom{0}0.15&0.18          & $0.75$ 	\\
NGC 1569  & 69.1  & \phantom{0}2.87  & $2.11$                 & 0.8 & 2.7 & $14.81$           	      & $\phantom{0}0.98$ 	& 0.05 & \phantom{0}4.11&0.02          & $2.39$ 		\\ 
NGC 2366  & 63.0  &\phantom{0}3.34  & $7.87$ 		 & 1.4 & 7.9 & $\phantom{0}6.69$  & $\phantom{0}8.50$ 	& 1.08 & \phantom{0}3.39 &0.43         & $1.47$ 	\\
NGC 3738  & 22.6  & \phantom{0}3.82 & $6.54$		 & 0.5 & 1.4 & $\phantom{0}7.65$  & $33.17$ 	                   & 5.07 & 33.86                     &5.18        & $2.95$ \\
UGC 8508  & 83.5  &\phantom{0}2.60  & $0.43$                  & 0.5 & 1.9 & $\phantom{0}1.91$   & $\phantom{0}0.83$ 	& 1.92 & \phantom{0}4.29 &0.99        & $3.80$ 		\\
WLM		 & 74.0   & \phantom{0}0.90  & $0.73$                 & 0.8 & 3.0 & $\phantom{0}6.45$  & $\phantom{0}0.79$ 	& 1.08 & \phantom{0}4.59 &0.63       & $0.96$ \\
Haro 29	 & 61.2   &\phantom{0}4.70  & $0.68$                  & 0.4 & 4.0 & $\phantom{0}5.94$   & $\phantom{0}1.20$ 	& 1.76 & \phantom{0}5.60 &0.82       & $7.58$\\
Haro 36	 & 70.0   & \phantom{0}8.91   & $3.27$		 & 0.6 & 3.0 & $10.44$                      & $\phantom{0}1.00$ 	& 0.31 & \phantom{0}6.26 &0.19        & $3.45$ 		\\
\hline
\end{tabular}
\label{t1}
Table Columns: Galaxy name, quoted inclination, NED Distance, Total B-Band Luminosity, Disk Scale Length, Distance to last data point, Total gas mass, Disk mass (CG), Mass to light ratio for the CG mass, Disk mass (MOND), Mass to light ratio for the MOND mass, scaled observed centripetal acceleration of the last data point.
\end{table*}

\end{document}